\documentclass[10pt]{article}
\usepackage{latexsym}
\usepackage{amssymb}
\usepackage{amsmath}
\usepackage{amscd}
\usepackage{amsthm}
\usepackage[left=2cm,top=2.5cm,right=2.5cm,bottom=1.5cm]{geometry}
\usepackage[dvips]{graphicx}
\usepackage{hyperref}
\usepackage{textcomp}

\begin{document}
\begin{center}
\Large{\bf{Some Invariant String Cosmological Models in Cylindrically Symmetric Space-time}}

\vspace{10mm}

\normalsize{Ahmad T Ali$^{1,\dag}$, Anil Kumar Yadav$^{2}$ ,
Farook Rahaman$^3$ and Arkopriya Mallick$^{4}$} \\

\vspace{4mm}
\normalsize{$^1$ King Abdul Aziz University,
Faculty of Science, Department of Mathematics,\\
PO Box 80203, Jeddah, 21589, Saudi Arabia.}\\
E-mail: atali71@yahoo.com\\
\normalsize{$^\dag$ Mathematics Department, Faculty of Science, Al-Azhar University,\\
Nasr city, 11884, Cairo, Egypt}\\
\vspace{2mm}
\normalsize{$^2$ Department of Physics, Anand Engineering College,
Keetham, Agra - 282 007, India.\\
E-mail: abanilyadav@yahoo.co.in}\\
\vspace{2mm}
\normalsize{$^3$ Department of Mathematics,
Jadavpur University, Kolkata 700 032, India. \\
E-mail: rahaman@iucaa.ernet.in}\\
\vspace{2mm} \normalsize{$^4$ Department of Mathematics,
Jadavpur University, Kolkata 700 032, India. \\
E-mail: arkopriyamallick@gmail.com}
\end{center}

\begin{abstract}
In this paper we derive some new invariant solutions of
Einstein-Maxwell's field equations for string fluid as source of
matter in cylindrically symmetric space-time  with Variable Magnetic
Permeability.  We also discuss the physical and geometrical
properties of the models
 derived in the paper. The solutions, at least one of them, are
  interesting physically as they can explain   the accelerating as
  well as  singularity free
  Universe.

\end{abstract}



\section{Introduction }

\vspace{10mm}

One of the important consequence  of general theory of relativity is
to know the large scale structure of the Universe. The study of
cosmology is devoted to the Universe as a whole. Initial study was
based on FRW type which describes the space which is homogeneous and
isotropic. We know homogeneity and isotropy are symmetries of space.
Though the present observations indicates that the Universe is
almost perfectly homogeneous and isotropic in large scales,,
however, the symmetries of space in very early Universe could be
very different. Today's observations could not able to provide
information  about symmetries of space near initial singularity.
Hence, it is very justified to consider inhomogeneous and
anisotropic models of the Universe. Cylindrically-symmetric
space-time is more general than the  homogeneous and isotropic
 space-time and plays an important role in the study of the universe
when the anisotropy and inhomogeneity are taking into consideration. Barrow and Kunze
 \cite{barr1, barr2} investigated flat and open homogeneous universe, considering string
 fluid as source of matter. In the recent past, Pradhan et al \cite{prad11} and Yadav et al
  \cite{yadav11} have studied inhomogeneous cosmological models with cloud of strings;
  the latter were invoked to palliate the problems associated with cylindrically symmetric space-time.

Due to non-linearity,  general theory of relativity is a very
difficult theory.   Therefore, to  describe the models of the
Universe, different group of symmetries are used in literature.
  In this study, we apply the so-called symmetry analysis
method.
 The main advantage of this method is that it can be successfully applied to nonlinear differential
 equations.
The similarity solutions are attractive  because they result in the
reduction of the independent variables of the problem.
  In our case, the problem under investigation is the system of second order nonlinear
  PDEs  and  similarity solution will transform this system of nonlinear PDEs into a system of ODEs.

The  groups of continuous transformations that leave a given family
of equations invariant is known as  symmetry groups (isovector fields) \cite{att1,ali1,mekh1,ali2,ali3}.
 In a pioneering work,
Ovsiannikov \cite{ovsi1} mentioned  that the usual Lie infinitesimal
invariance approach could as well be employed in order to construct
symmetry groups \cite{blum1,ibra1,olve1}.

In  recent  years , there has been  considerable  interest  in
string  cosmology. Cosmic  string  plays  an  important  role  in
the  study  of  the  early  universe.  String  cosmology  is  a
relatively  new  field  that  tries  to  apply  equation  of  string
theory  to  solve   the  questions  of  early  cosmology , a related
area  of  study  in  brain  cosmology. It was argued by Melvin
\cite{mel1}   that the presence of magnetic field is not as
unrealistic    because for a large part of the history of evolution
matter was highly ionized and matter and field were smoothly
coupled. Latter during the expansion of the Universe, the ions
combined to form neutral matter. Therefore, inclusion of the
magnetic field with Variable Magnetic Permeability is justified for
cosmological  modeling of the Universe.

In this paper, we attempted to find a new class of exact
(similarity) solutions for string   cosmological  models  in
cylindrically  symmetric  inhomogeneous  universe  with  electro
magnetic   perfect  fluid  distribution  with  variable  magnetic
permeability  in  general  relativity.

We organize the paper   as follows:
 String cosmological model in cylindrically symmetric inhomogeneous universe with electro-magnetic
 perfect fluid distribution
of variable magnetic permeability in general relativity, is
introduced in section 2. In section 3,  symmetry analysis and
isovector fields for Einstein field equations are obtained. In
section 4,  we find a new class of exact (similarity) solutions for
Einstein field equations. Section 5 is devoted to  the study of some
physical and geometrical properties of the obtained model. The paper
ends with a short discussion.

\section{The metric and field equations}

We consider the metric in the form
\begin{equation}  \label{u21}
ds^2=A^2\,\big(dx^2-dt^2\big)+B^2\,dy^2+C^2\,dz^2,
\end{equation}
where $A$, $B$ and $C$ are functions of $x$ and $t$. The energy-momentum tensor for the string with electro-magnetic field has the form
\begin{equation}  \label{u22}
T_{ij}=\rho\,u_i\,u_j-\lambda\,x_i\,x_j+E_{ij},
\end{equation}
where $u_i$ and $x_i$ satisfy the conditions
\begin{equation}  \label{u23}
u^i\,u_i\,=\,-x^i\,x_i\,=\,-1,
\end{equation}
and
\begin{equation}  \label{u24}
u^i\,x_i\,=\,0.
\end{equation}
Here $\rho$ being the rest energy density of the system of strings, $\lambda$ the tension density of the strings , $x^i$ is a unit space-like vector representing the direction of strings so that $x^1=x^2=x^4=0$ and $x^3\neq0$, and $u^i$ is the four velocity vector. $E_{ij}$ is the electro-magnetic field given by Lichnerowicz \cite{lich1}:
\begin{equation}  \label{u25}
E_{ij}=\bar{\mu}\big[h_lh^l\big(u_iu_j+\frac{1}{2}g_{ij}\big)-h_ih_j\big],
\end{equation}
where $\bar{\mu}$ is the magnetic permeability and $h_i$ the magnetic flux vector defined by:
\begin{equation}  \label{u26}
h_i=\dfrac{1}{\bar{\mu}}\,^*F_{ji}\,u^j,
\end{equation}
where the dual electro-magnetic field tensor $^*F_{ij}$ is defined by Synge \cite{synge1}
\begin{equation}  \label{u27}
^*F_{ij}=\dfrac{\sqrt{-g}}{2}\,\epsilon_{ijkl}\,F^{kl}.
\end{equation}
Here $F_{ij}$ is the electro-magnetic field tensor and $\epsilon_{ijkl}$ is a Levi-Civita tensor density. In the present scenario, the co-moving coordinates are taken as
\begin{equation}  \label{u28}
u^i=\Big(0,0,0,\dfrac{1}{A}\Big).
\end{equation}
We choose the direction of string parallel to $x$-axis so that
\begin{equation}  \label{u29}
x^i=\Big(\dfrac{1}{A},0,0,0\Big).
\end{equation}
If we consider the current flow along $z$-axis, then $F_{12}$ is only non-vanishing component of $F_{ij}$. Then the Maxwell's equations
\begin{equation}  \label{u28}
F_{ij;k}+F_{jk;i}+F_{ki;j}=0
\end{equation}
and
\begin{equation}  \label{u29}
\Big[\dfrac{1}{\bar{\mu}}F^{ij}\Big]_{;j}=J^i
\end{equation}
require that $F_{12}$ be function of $x$ alone \cite{prad12}. We assume that the magnetic permeability as
 a function of both $x$ and $t$. Here the semicolon represents a covariant differentiation.\\

The Einstein's field equation
\begin{equation}  \label{u29-0}
R_{ij}-\dfrac{1}{2}\,g_{ij}\,R=-\chi\,T_{ij},
\end{equation}
for the line-element (\ref{u21}) lead to the following system of equations:

\begin{equation}  \label{u210}
\begin{array}{ll}
    \dfrac{C_{xt}}{C}+\dfrac{B_{xt}}{B}-\dfrac{A_t\,C_x+A_x\,C_t}{A\,C}-\dfrac{A_t\,B_x+A_x\,B_t}{A\,B}=0,
  \end{array}
\end{equation}

\begin{equation}  \label{u211}
\begin{array}{ll}
        \dfrac{C_{xx}-C_{tt}}{2\,C}+\dfrac{B_{xx}-B_{tt}}{2\,B}+\dfrac{A_{xx}-A_{tt}}{A}+\dfrac{A_{t}^2-A_{x}^2}{A^2}=0,
  \end{array}
\end{equation}

\begin{equation}  \label{u212}
\begin{array}{ll}
        \chi\,A^2\,\lambda=\dfrac{C_{xx}}{C}+\dfrac{B_{tt}}{B}+\dfrac{A_{xx}-A_{tt}}{A}+\dfrac{B_t\,C_t-B_x\,C_x}{B\,C}\\
        \\
        \,\,\,\,\,\,\,\,\,\,\,\,\,\,\,\,\,\,\,\,\,\,\,\,\,\,\,\,\,\,\,\,\,\,\,\,\,\,\,\,\,\,\,\,\,\,\,\,\,\,
        -\dfrac{A_t\,C_t+A_x\,C_x}{A\,C}-\dfrac{A_x\,B_x+A_t\,B_t}{A\,B}+\dfrac{A_t^2-A_x^2}{A^2},
  \end{array}
\end{equation}

\begin{equation}  \label{u213}
\begin{array}{ll}
        \chi\,A^2\,\rho=\dfrac{C_{tt}-2\,C_{xx}}{C}-\dfrac{B_{tt}}{B}+\dfrac{A_{tt}-A_{xx}}{A}+\dfrac{B_t\,C_t-B_x\,C_x}{B\,C}\\
        \\
        \,\,\,\,\,\,\,\,\,\,\,\,\,\,\,\,\,\,\,\,\,\,\,\,\,\,\,\,\,\,\,\,\,\,\,\,\,\,\,\,\,\,\,\,\,\,\,\,\,\,
        +\dfrac{A_t\,C_t+A_x\,C_x}{A\,C}+\dfrac{A_x\,B_x+A_t\,B_t}{A\,B}+\dfrac{A_x^2-A_t^2}{A^2},
  \end{array}
\end{equation}

\begin{equation}  \label{u214}
\begin{array}{ll}
        \dfrac{\chi\,F_{12}^2}{2\,\bar{\mu}\,B^2}=\dfrac{C_{xx}-C_{tt}}{C}+\dfrac{A_{xx}-A_{tt}}{A}+\dfrac{A_t^2-A_x^2}{A^2},
  \end{array}
\end{equation}

The velocity field $u^i$ is ir-rotational. The scalar expansion
$\Theta$, shear scalar $\sigma^2$, acceleration vector $\dot{u}_i$
and Proper volume $V$ are respectively found to have the following
expressions \cite{dec1,rayc1}:

\begin{equation}  \label{u215}
\Theta\,=\,u_{;i}^{i}=\dfrac{1}{A}\Big(\dfrac{C_t}{C}+\dfrac{B_t}{B}+\dfrac{A_t}{A}\Big),
\end{equation}

\begin{equation}  \label{u216}
\begin{array}{ll}
\sigma^2\,=\,\dfrac{1}{2}\,\sigma_{ij}\,\sigma^{ij}=\dfrac{\Theta^2}{3}-\dfrac{1}{A^2}\Big(
\dfrac{B_t\,C_t}{B\,C}+\dfrac{A_t\,C_t}{A\,C}+\dfrac{A_t\,B_t}{A\,B}\Big),
\end{array}
\end{equation}

\begin{equation}  \label{u217}
\dot{u}_i\,=\,u_{i;j}\,u^j\,=\,\Big(\dfrac{A_x}{A},0,0,0\Big),
\end{equation}

\begin{equation}  \label{u218}
V=\sqrt{-g}=A^2\,B\,C,
\end{equation}
where $g$ is the determinant of the metric (\ref{u21}). The shear tensor is
\begin{equation}  \label{u219}
  \begin{array}{ll}
\sigma_{ij}\,=\,u_{(i;j)}+\dot{u}_{(i}\,u_{j)}-\frac{1}{3}\,\Theta\,(g_{ij}+u_i\,u_j).
\end{array}
\end{equation}
and the non-vanishing components of the $\sigma_i^j$ are
\begin{equation}  \label{u220}
\left\{
  \begin{array}{ll}
    \sigma_1^1\,&=\,\dfrac{1}{3\,A}\Big(\dfrac{2\,A_t}{A}-\dfrac{B_t}{B}-\dfrac{C_t}{C}\Big),\\
    \\
\sigma_2^2\,&=\,\dfrac{1}{3\,A}\Big(\dfrac{2\,B_t}{B}-\dfrac{C_t}{C}+\dfrac{A_t}{A}\Big),\\
\\
\sigma_3^3\,&=\,\dfrac{1}{3\,A}\Big(\dfrac{2\,C_t}{C}-\dfrac{B_t}{B}+\dfrac{A_t}{A}\Big),\\
\\
\sigma_4^4\,&=0.
   \end{array}
\right.
\end{equation}
Using the field equations and the relations (\ref{u215}) and (\ref{u216}) one obtains the Raychaudhuri's equation is
\begin{equation}  \label{u221}
  \begin{array}{ll}
\dfrac{\partial\Theta}{\partial t}=\dot{u}_{;i}^i-\dfrac{\Theta^2}{3}-2\,\sigma^2-\dfrac{\rho_p}{2},
   \end{array}
\end{equation}
where
\begin{equation}  \label{u222}
  \begin{array}{ll}
\rho_p=2\,R_{ij}\,u^i\,u^j.
   \end{array}
\end{equation}

The Einstein field equations (\ref{u210})-(\ref{u214}) constitute a system of five highly non-linear differential equations with six unknowns variables, $A$, $B$, $C$, $\lambda$, $\rho$ and $\frac{F_{12}^2}{\bar{\mu}}$. Therefore, one physically reasonable conditions amongst these parameters are required to obtain explicit solutions of the field equations. Let us assume that the expansion scalar $\Theta$ in the model (\ref{u21}) is proportional to the eigenvalue $\sigma_1^1$ of the shear tensor $\sigma_j^k$. Then from (\ref{u215}) and (\ref{u220}), we get
\begin{equation}\label{u223}
  \begin{array}{ll}
\dfrac{2\,A_t}{A}-\dfrac{B_t}{B}-\dfrac{C_t}{C}=3\,\gamma\,\Big(\dfrac{A_t}{A}+\dfrac{B_t}{B}+\dfrac{C_t}{C}\Big),
  \end{array}
\end{equation}
where $\gamma$ is a constant of proportionality. The above equation can be written in the form
\begin{equation}\label{u224}
  \begin{array}{ll}
\dfrac{A_t}{A}\,=n\,\Big(\dfrac{B_t}{B}+\dfrac{C_t}{C}\Big).
  \end{array}
\end{equation}
where $n=\dfrac{1+3\,\gamma}{2-3\,\gamma}$. If we integrate the above equation with respect to $t$, we can get the following relation
\begin{equation}\label{u225}
  \begin{array}{ll}
A(x,t)\,=\,f(x)\,\Big(B(x,t)\,C(x,t)\Big)^n,
  \end{array}
\end{equation}
where  $f(x)$ is a constant of integration which is an arbitrary function of $x$. If we substitute the metric function $A$ from (\ref{u217}) in the Einstein field equations, the equations (\ref{u210})-(\ref{u211}) transform to the nonlinear partial differential equations of the coefficients $B$ and $C$ only, as the following new form:
\begin{equation}  \label{u210-1}
\begin{array}{ll}
    E_1=\dfrac{B_{xt}}{B}+\dfrac{C_{xt}}{C}-2\,n\,\Big(\dfrac{B_x\,B_t}{B^2}+\dfrac{B_x\,C_t+B_t\,C_x}{B\,C}
    +\dfrac{C_x\,C_t}{C^2}\Big)-\dfrac{f'}{f}\Big(\dfrac{B_t}{B}+\dfrac{C_t}{C}\Big)=0,
  \end{array}
\end{equation}

\begin{equation}  \label{u211-1}
\begin{array}{ll}
        E_2=\Big(n+\dfrac{1}{2}\Big)\Big[\dfrac{B_{xx}-B_{tt}}{B}+\dfrac{C_{xx}-C_{tt}}{C}\Big]
        +n\Big(\dfrac{B_t^2-B_x^2}{B^2}+\dfrac{C_t^2-C_x^2}{C^2}\Big)+\dfrac{f''}{f}-\dfrac{f'^2}{f^2}=0,
  \end{array}
\end{equation}
where the prime indicates derivative with respect to the coordinate $x$.

\section{Symmetry analysis method}

In order to obtain an exact solutions of the system of nonlinear partial differential equations (\ref{u210-1})-(\ref{u211-1}), we will use the symmetry
analysis method. For this we write
\begin{equation}\label{u31}
\left\{
\begin{array}{ll}
x_i^{*}=x_i+\epsilon\,\xi_{i}(x_j,u_{\beta})+\bold{o}(\epsilon^2),\\
u_{\alpha}^{*}=u_{\alpha}+\epsilon\,\eta_{\alpha}(x_j,u_{\beta})+\bold{o}(\epsilon^2),
\end{array}
\right.
\,\,\,i,j,\alpha,\beta=1,2,
\end{equation}
as the infinitesimal Lie point transformations. We have assumed
that the system (\ref{u210-1})-(\ref{u211-1}) is invariant under the transformations given in
Eq. (\ref{u31}). The corresponding infinitesimal generator of Lie groups
(symmetries) is given by
\begin{equation}\label{u32}
 X=\sum_{i=1}^{2}\xi_{i}\dfrac{\partial}{\partial x_{i}}+\sum_{\alpha=1}^{2}\eta_{\alpha}
 \dfrac{\partial}{\partial u_{\alpha}},
 \end{equation}
where $x_1=x$, $x_2=t$, $u_1=B$ and $u_2=C$. The coefficients $\xi_{1}$, $\xi_{2}$, $\eta_{1}$ and $\eta_{2}$ are the functions of $x$, $t$, $B$ and $C$.
These coefficients are the components of infinitesimals symmetries
corresponding to $x$, $t$, $B$ and $C$ respectively, to be determined from the invariance conditions:
\begin{equation}\label{u33}
{\text{Pr}}^{(2)}\,X\Big(E_m\Big)|_{E_m=0}=0,
\end{equation}
where $E_m=0,\,m=1,2$ are the system (\ref{u210-1})-(\ref{u211-1}) under study and
${\text{Pr}}^{(2)}$ is the second prolongation of the symmetries $X$.
Since our equations (\ref{u210-1})-(\ref{u211-1}) are at most of order two, therefore, we
need second order prolongation of the infinitesimal generator
in Eq. (\ref{u33}). It is worth noting that, the $2$-th order prolongation is given by:
\begin{equation}\label{u34}
{\text{Pr}}^{(2)}\,X=\sum_{i=1}^{2}\xi_{i}\dfrac{\partial}{\partial x_{i}}+\sum_{\alpha=1}^{2}\eta_{\alpha}
 \dfrac{\partial}{\partial u_{\alpha}}+\sum_{i=1}^{2}\,\sum_{\alpha=1}^{2}\,\eta_{\alpha\,i}\,\dfrac{\partial}{\partial u_{\alpha,i}}
+\sum_{j=1}^{2}\,\sum_{i=1}^{2}\,\sum_{\alpha=1}^{2}\,\eta_{\alpha\,i\,j}\,\dfrac{\partial}{\partial u_{\alpha,ij}},
\end{equation}
where
\begin{equation}\label{u35}
\eta_{\alpha\,i}=D_{i}\Big(\eta_{\alpha}\Big)-\sum_{j=1}^{2}\,u_{\alpha,j}\,D_i\Big(\xi_j\Big)\,,\,\,\,\,\,\,\,
\eta_{\alpha\,i\,j}=D_{j}\Big(\eta_{\alpha\,i}\Big)-\sum_{k=1}^{2}\,u_{\alpha,k\,i}\,D_j\Big(\xi_k\Big)\,.
\end{equation}
The operator $D_{i}$ is called the {\it total derivative} ({\it Hach operator}) and taken the following
form:
\begin{equation}\label{u36}
D_i=\dfrac{\partial}{\partial x_i}+\sum_{\alpha=1}^{2}\,u_{\alpha,i}\,\dfrac{\partial}{\partial u_{\alpha}}
+\sum_{j=1}^{2}\,\sum_{\alpha=1}^{2}\,u_{\alpha,j\,i}\,
\dfrac{\partial}{\partial u_{\alpha,j}},
\end{equation}
where $u_{\alpha,i}=\frac{\partial u_{\alpha}}{\partial x_{i}}$ and $u_{\alpha,i\,j}=\frac{\partial^2 u_{\alpha}}{\partial x_{j}\,\partial x_{i}}$.

Expanding the system of Eqs. (\ref{u33}) along with the original system of Eqs. (\ref{u210-1})-(\ref{u211-1}) to eliminate $B_{xx}$ and $B_{xt}$ while we set the coefficients involving $C_{x}$, $C_{t}$, $C_{xx}$, $C_{xt}$, $C_{tt}$,
$B_{x}$, $B_{t}$, $B_{tt}$ and
various products to zero give rise the essential set of over-determined
equations. Solving the set of these determining equations, the components of symmetries takes the following form:
\begin{equation}\label{u37}
\xi_{1}=c_1\,x+c_2,\,\,\,\,\,\xi_{2}=c_1\,t+c_3,\,\,\,\,\,\eta_{1}=c_4\,B ,\,\,\,\,\,\eta_{2}=c_5\,C,
\end{equation}
such that the function $A(t)$ must be equal:
\begin{equation}\label{u37-1}\left\{
                               \begin{array}{ll}
                                 f(x)=c_6\,\exp\big[c_7\,x\big],\,\,\,\,\,\,\,\,\,\,\,\mathrm{if}\,\,\, c_1=0, \\
\\
                                 f(x)=c_8\big(c_1\,x+c_2\big)^{c_9},\,\,\,\,\,\,\mathrm{if}\,\,\, c_1\neq0,
                               \end{array}
                             \right.
\end{equation}
where $c_i,\,i=1,2,...,9$ are an arbitrary constants.

\section{Similarity solutions}

The characteristic equations corresponding to the symmetries (\ref{u37}) are given by:
\begin{equation}\label{u41}
\dfrac{dx}{c_1\,x+c_2}=\dfrac{dt}{c_1\,t+c_3}=\dfrac{dB}{c_4\,B}=\dfrac{dC}{c_5\,C}.
\end{equation}
By solving the above system, we have the following two cases:\\

\textbf{Case (1):} When $c_1=0$, the similarity variable and similarity functions can be written as the following:
\begin{equation}\label{u42-1}
\begin{array}{ll}
\xi=a\,x+b\,t,\,\,\,\,\,\,B(x,t)=\Psi(\xi)\,\exp[c\,x],\,\,\,\,\,\,C(x,t)=\Phi(\xi)\,\exp[d\,x],
\end{array}
\end{equation}
where $a=c_3$, $b=-c_2$, $c=\dfrac{c_4}{c_2}$ and $d=\dfrac{c_5}{c_2}$ are an arbitrary constants. Substituting the transformations (\ref{u42-1}) in the field Eqs. (\ref{u217})-(\ref{u218}) lead
to the following system of ordinary differential equations:
\begin{equation}\label{u42-2}
\begin{array}{ll}
\dfrac{a\,\Psi''+\big[c_7-c+2\,n\,(c+d)\big]\,\Psi'}{\Psi}+\dfrac{a\,\Phi''+\big[c_7-d+2\,n\,(c+d)\big]\,\Phi'}{\Phi}
=2\,a\,n\Big(\dfrac{\Psi'}{\Psi}+\dfrac{\Phi'}{\Phi}\Big)^2,
\end{array}
\end{equation}

\begin{equation}\label{u42-3}
\begin{array}{ll}
(b^2-a^2)\,\Bigg[(2\,n+1)\,\Big(\dfrac{\Psi''}{\Psi}+\dfrac{\Phi''}{\Phi}\Big)
-2\,n\,\Big(\dfrac{\Psi'^2}{\Psi^2}+\dfrac{\Phi'^2}{\Phi^2}\Big)\Bigg]-2\,a\,\Big(\dfrac{c\,\Psi'}{\Psi}+\dfrac{d\,\Phi'}{\Phi}\Big)=c^2+d^2,
\end{array}
\end{equation}
The equations (\ref{u42-2}) and (\ref{u42-3}) are non-linear ordinary differential equations which is very difficult to solve. However, it is worth noting that, this equation is easy to solve when $b=a$. In this case, the equation (\ref{u42-3}) takes the form:
\begin{equation}\label{u42-4}
\begin{array}{ll}
\dfrac{c\,\Psi'}{\Psi}+\dfrac{d\,\Phi'}{\Phi}=-\dfrac{c^2+d^2}{2\,a}.
\end{array}
\end{equation}
By integration the above equation, we can get the following:
\begin{equation}\label{u42-6}
\begin{array}{ll}
\Phi(\xi)=q_1\,\Psi^{\alpha_1}(\xi)\,\exp\big[\alpha_2\,\xi\big],
\end{array}
\end{equation}
where $\alpha_1=-\dfrac{c}{d}$ and $\alpha_2=-\Big(\dfrac{c^2+d^2}{2\,a\,d}\Big)$ while $q_1$ is an arbitrary constant of integration. Substitute (\ref{u42-6}) in (\ref{u42-2}), we have the following ordinary differential equation of the function $\Psi$ only as follows:

\begin{equation}\label{u42-7}
\begin{array}{ll}
\dfrac{\Psi''}{\Psi}=(\alpha_3-1)\,\Big(\dfrac{\Psi'^2}{\Psi^2}\Big)
+\alpha_4\,\Big(\dfrac{\Psi'}{\Psi}\Big)
+\alpha_5,
\end{array}
\end{equation}
where
\begin{equation}\label{u42-7-1}\left\{
                               \begin{array}{ll}
\alpha_3=-\dfrac{1+\alpha_1^2-2\,n\,(1+\alpha_1)^2}{1+\alpha_1},\\
\\
\alpha_4=-\dfrac{2\,\alpha_2\,\Big[a_7\,(1+\alpha_1)+d\,\alpha_1\,\Big(1+\alpha_1^2-2\,n\,(1+\alpha_1)^2\Big)\Big]}{d\,(1+\alpha_1)\,(1+\alpha_1^2)},\\
\\
\alpha_5=-\dfrac{\alpha_2^2\,\Big[2\,a_7+d\,\Big(\alpha_1^2-1-2\,n\,(\alpha_1^2+2\,\alpha_1-1)\Big)\Big]}{d\,(1+\alpha_1)\,(1+\alpha_1^2)}.
                               \end{array}
                             \right.
\end{equation}
If we use the transformation
\begin{equation}\label{u42-8}
                               \begin{array}{ll}
\Psi(\xi)\,=q_2\,\exp\Big[\int\,\Omega(\xi)\,d\xi\Big]
                               \end{array}
\end{equation}
the equation (\ref{u42-7}) becomes:
\begin{equation}\label{u42-9}
\begin{array}{ll}
\Omega'=\alpha_3\,\Omega^2+\alpha_4\,\Omega+\alpha_5,
\end{array}
\end{equation}
where $q_2$ is constant while $\Omega(\xi)$ is a new function of $\xi$. For solving the above ordinary differential equation, we must take the following cases:\\

\textbf{Case (1.1):} When $\alpha_3\,\neq\,0$, $\alpha_4\,\neq\,0$ and $\alpha_5\,\neq\,0$, there exists three cases as the following:\\

\textbf{Case (1.1.1):} When $4\,\alpha_3\,\alpha_5-\alpha_4^2=\dfrac{4K_1^2}{a^2}$, the general solution of the equation (\ref{u42-9}) is:
\begin{equation}\label{u42-10}
\begin{array}{ll}
\Omega(\xi)=-\dfrac{\alpha_4}{2\,\alpha_3}-\dfrac{K_1}{a\,\alpha_3}\,\mathrm{tan}\Big[\dfrac{K_1\,\xi}{a}+\xi_0\Big],
\end{array}
\end{equation}
where $\xi_0$ is an arbitrary constant of integration.\\

\textbf{Case (1.1.2):} When $\alpha_4^2-4\,\alpha_3\,\alpha_5=\dfrac{4K_2^2}{a^2}$, the general solution of the equation (\ref{u42-9}) is:
\begin{equation}\label{u42-11}
\begin{array}{ll}
\Omega(\xi)=-\dfrac{\alpha_4}{2\,\alpha_3}+\dfrac{K_2}{a\,\alpha_3}\,\mathrm{tanh}\Big[\dfrac{K_2\,\xi}{a}+\xi_0\Big],
\end{array}
\end{equation}
where $\xi_0$ is an arbitrary constant of integration.\\

\textbf{Case (1.1.3):} When $\alpha_4^2=4\,\alpha_3\,\alpha_5$, the general solution of the equation (\ref{u42-9}) is:
\begin{equation}\label{u42-12}
\begin{array}{ll}
\Omega(\xi)=\dfrac{\alpha_4\,\xi-2}{2\,\alpha_3\,\xi}.
\end{array}
\end{equation}

\textbf{Case (1.2):} When $\alpha_3\,\neq\,0$, $\alpha_4\,\neq\,0$ and $\alpha_5\,=\,0$, the general solution of the equation (\ref{u42-9}) is:
\begin{equation}\label{u42-13}
\begin{array}{ll}
\Omega(\xi)=\dfrac{\alpha_4}{\exp\Big[\alpha_3-\alpha_4\,\big(\xi+\xi_0\big)\Big]},
\end{array}
\end{equation}
where $\xi_0$ is an arbitrary constant of integration.\\

\textbf{Case (1.3):} When $\alpha_3\,\neq\,0$, $\alpha_4\,=\,0$ and $\alpha_5\,\neq\,0$, then there exists the following cases:\\

\textbf{Case (1.3.1):} When $\alpha_3\,=\,\pm K_4^2$ and $\alpha_5\,=\,\pm K_3^2$, the general solution of the equation (\ref{u42-9}) is:
\begin{equation}\label{u42-14-1}
\begin{array}{ll}
\Omega(\xi)=\pm\dfrac{K_3}{K_4}\,\mathrm{tan}\Big[K_4\,K_3\,\big(\xi+\xi_0\big)\Big],
\end{array}
\end{equation}
where $\xi_0$ is an arbitrary constant of integration.\\

\textbf{Case (1.3.2):} When $\alpha_3\,=\,\mp K_6^2$ and $\alpha_5\,=\,\pm K_5^2$, the general solution of the equation (\ref{u42-9}) is:
\begin{equation}\label{u42-14-2}
\begin{array}{ll}
\Omega(\xi)=\pm \dfrac{K_5}{K_6}\,\mathrm{tanh}\Big[K_5\,K_6\,\big(\xi+\xi_0\big)\Big],
\end{array}
\end{equation}
where $\xi_0$ is an arbitrary constant of integration.\\

\textbf{Remark (1):} The two cases (1.3.1) and (1.3.2) are the spacial cases from the two cases (1.1.1) and (1.1.2), respectively.\\

\textbf{Case (1.4):} When $\alpha_3\,=\,0$, $\alpha_4\,\neq\,0$ and $\alpha_5\,\neq\,0$, the general solution of the equation (\ref{u42-9}) is:
\begin{equation}\label{u42-15}
\begin{array}{ll}
\Omega(\xi)=\exp\Big[\alpha_4\big(\xi+\xi_0\big)\Big]-\dfrac{\alpha_5}{\alpha_4},
\end{array}
\end{equation}
where $\xi_0$ is an arbitrary constant of integration.\\

\textbf{Case (1.5):} When $\alpha_3\,=\,\alpha_4\,=\,0$ and $\alpha_5\,\neq\,0$, the general solution of the equation (\ref{u42-9}) is:
\begin{equation}\label{u42-16}
\begin{array}{ll}
\Omega(\xi)=\alpha_5\,\big(\xi+\xi_0\big),
\end{array}
\end{equation}
where $\xi_0$ is an arbitrary constant of integration.\\

\textbf{Case (1.6):} When $\alpha_3\,=\,\alpha_5\,=\,0$ and $\alpha_4\,\neq\,0$, the general solution of the equation (\ref{u42-9}) is:
\begin{equation}\label{u42-17}
\begin{array}{ll}
\Omega(\xi)=\exp\Big[\alpha_4\,\big(\xi+\xi_0\big)\Big],
\end{array}
\end{equation}
where $\xi_0$ is an arbitrary constant of integration.\\

\textbf{Case (1.7):} When $\alpha_4\,=\,\alpha_5\,=\,0$ and $\alpha_3\,\neq\,0$, the general solution of the equation (\ref{u42-9}) is:
\begin{equation}\label{u42-18}
\begin{array}{ll}
\Omega(\xi)=-\dfrac{1}{\alpha_3\,\big(\xi+\xi_0\big)},
\end{array}
\end{equation}
where $\xi_0$ is an arbitrary constant of integration.\\

\textbf{Case (1.8):} When $\alpha_3\,=\,\alpha_4\,=\,\alpha_5\,=\,0$, the general solution of the equation (\ref{u42-9}) is:
\begin{equation}\label{u42-19}
\begin{array}{ll}
\Omega(\xi)=\xi_0,
\end{array}
\end{equation}
where $\xi_0$ is an arbitrary constant of integration.\\

\textbf{Remark (2):} The three cases (1.6), (1.7) and (1.8) are the spacial cases from the three cases (1.4), (1.1.3) and (1.5), respectively.\\

Every solution from above give the solution of Einstein field equation. Therefore, we will study some of these solutions as the following:\\

\textbf{Solution (1.1.1):} We consider the solution correspondence to the case (1.1.1). Without loss of generality, we can take, in this case, $d=-c$. Now, by using (\ref{u42-10}),  (\ref{u42-8}), (\ref{u42-6}) and (\ref{u42-1}), we can find the solution as the following:
\begin{equation}\label{uu1}
\left\{
  \begin{array}{ll}
A(x,t)=q_1\,\exp\Big[c_7\Big(x+n\,n_1(x+t)\Big)\Big]\,\cos^{2\,n\,n_1}\Big[K_1\,\big(x+t\big)+\xi_0\Big],\\
\\
B(x,t)=q_2\,\exp\Big[\dfrac{c}{2}\Big(x-t+n_2(x+t)\Big)\Big]\,\cos^{n_1}\Big[K_1\,\big(x+t\big)+\xi_0\Big],\\
\\
C(x,t)=q_3\,\exp\Big[\dfrac{c}{2}\Big(t-x+n_2(x+t)\Big)\Big]\,\cos^{n_1}\Big[K_1\,\big(x+t\big)+\xi_0\Big],
  \end{array}
\right.
\end{equation}
where $n=\dfrac{c^2+c_7^2+4\,K_1^2}{4\,c^2}$, $n_1=-\dfrac{c^2}{c_7^2+4\,K_1^2}$ and $n_2=-\dfrac{c\,c_7}{c_7^2+4\,K_1^2}$. It is observed from equations (\ref{uu1}), the line element (\ref{u21}) can be written in the following form:
\begin{equation}  \label{s111}
\begin{array}{ll}
ds_{111}^2=q_1^2\,\exp\Big[2\,c_7\Big(x+n\,n_1\,(x+t)\Big)\Big]\,\cos^{4\,n\,n_1}\big[\xi\big]\,\Big(dx^2-dt^2\Big)
\\
\\
\,\,\,
+\exp\Big[c\,\Big(x-t+n_2\,(x+t)\Big)\Big]\,\cos^{2\,n_1}\big[\xi\big]\,
\Big(q_2^2\,dy^2+q_3^2\,\exp\Big[2\,c\,\big(t-x)\big)\Big]\,dz^2\Big).
\end{array}
\end{equation}
where $\xi=K_1\,\big(x+t\big)+\xi_0$ while $q_1$, $q_2$, $q_3$, $c$, $c_7$, $K_1$ and $\xi_0$ are an arbitrary constants.\\

\textbf{Solution (1.1.2):} We consider the solution correspondence the case (1.1.2). Without loss of generality, we can take, in this case, $d=-c$. Now, by using (\ref{u42-11}),  (\ref{u42-8}), (\ref{u42-6}) and (\ref{u42-1}), we can find the solution as the following:
\begin{equation}\label{uu2}
\left\{
  \begin{array}{ll}
A(x,t)=q_1\,\exp\Big[c_7\Big(x+n\,n_1(x+t)\Big)\Big]\,\cosh^{2\,n\,n_1}\Big[K_2\,\big(x+t\big)+\xi_0\Big],\\
\\
B(x,t)=q_2\,\exp\Big[\dfrac{c}{2}\Big(x-t+n_2(x+t)\Big)\Big]\,\cosh^{n_1}\Big[K_2\,\big(x+t\big)+\xi_0\Big],\\
\\
C(x,t)=q_3\,\exp\Big[\dfrac{c}{2}\Big(t-x+n_2(x+t)\Big)\Big]\,\cosh^{n_1}\Big[K_2\,\big(x+t\big)+\xi_0\Big],
  \end{array}
\right.
\end{equation}
where $n=\dfrac{c^2+c_7^2-4\,K_2^2}{4\,c^2}$, $n_1=-\dfrac{c^2}{c_7^2-4\,K_2^2}$ and $n_2=-\dfrac{c\,c_7}{c_7^2-4\,K_2^2}$. It is observed from equations (\ref{uu2}), the line element (\ref{u21}) can be written in the following form:
\begin{equation}  \label{s112}
\begin{array}{ll}
ds_{112}^2=q_1^2\,\exp\Big[2\,c_7\Big(x+n\,n_1\,(x+t)\Big)\Big]\,\cosh^{4\,n\,n_1}\big[\xi\big]\,\Big(dx^2-dt^2\Big)
\\
\\
\,\,\,
+\exp\Big[c\,\Big(x-t+n_2\,(x+t)\Big)\Big]\,\cosh^{2\,n_1}\big[\xi\big]\,
\Big(q_2^2\,dy^2+q_3^2\,\exp\Big[2\,c\,\big(t-x)\big)\Big]\,dz^2\Big).
\end{array}
\end{equation}
where $\xi=K_2\,\big(x+t\big)+\xi_0$ while $q_1$, $q_2$, $q_3$, $c$, $c_7$, $K_1$ and $\xi_0$ are an arbitrary constants.\\

\textbf{Solution (1.1.3):} We consider the solution correspondence the case (1.1.3). Without loss of generality, we can take, in this case, $d=-c$ and $a=1$. Now, by using (\ref{u42-12}),  (\ref{u42-8}), (\ref{u42-6}) and (\ref{u42-1}), we can find the solution as the following:
\begin{equation}\label{uu3}
\left\{
  \begin{array}{ll}
A(x,t)=q_1\,(x+t)^{-2\,n\,d_2^2}\exp\Big[2\,d_1\,d_2\Big(n\,t+n_1\,x\Big)\Big],\\
\\
B(x,t)=q_2\,(x+t)^{-d_2^2}\exp\Big[n_+\,t+n_-\,x\Big)\Big],\\
\\
C(x,t)=q_1\,(x+t)^{-d_2^2}\exp\Big[n_-\,t+n_+\,x\Big)\Big],
  \end{array}
\right.
\end{equation}
where $n=\dfrac{d_2^2+1}{4\,d_2^2}$, $n_1=\dfrac{d_2^2-3}{4\,d_2^2}$ and $n_{\pm}=d_1\,(d_2\pm1)$. It is observed from equations (\ref{uu3}), the line element (\ref{u21}) can be written in the following form:

\begin{equation}  \label{s113}
\begin{array}{ll}
ds_{113}^2=q_1^2\,(x+t)^{-4\,n\,d_2^2}\exp\Big[4\,d_1\,d_2\Big(n\,t+n_1\,x\Big)\Big]\,\Big(dx^2-dt^2\Big)
\\
\\
\,\,\,
+(x+t)^{-2\,d_2^2}\,
\Big(q_2^2\,\exp\Big[2\,(n_+\,t+n_-\,x)\Big]\,dy^2+q_3^2\,\exp\Big[2(n_-\,t+n_+\,x)\Big]\,dz^2\Big),
\end{array}
\end{equation}
where $q_1$, $q_2$, $q_3$, $d_1$ and $d_2$ are an arbitrary constants.\\

\textbf{Solution (1.2):} We consider the solution correspondence the case (1.2). Without loss of generality, we can take, in this case, $d=-c$, $c_7=2\,n\,c$, $n=\dfrac{1}{5}$ and $a=-5\,a_1$. Now, by using (\ref{u42-13}),  (\ref{u42-8}), (\ref{u42-6}) and (\ref{u42-1}), we can find the solution as the following:
\begin{equation}\label{uu4}
\left\{
  \begin{array}{ll}
A(x,t)=q_1\,\Big[a_1-\mathrm{e}^{\dfrac{3\,c\,(x+t)}{5}}\Big]^2\,\mathrm{e}^{-\dfrac{c}{5}\,(5\,t+7\,x)},\\
\\
B(x,t)=q_2\,\Big[a_1-\mathrm{e}^{\dfrac{3\,c\,(x+t)}{5}}\Big]^5\,\mathrm{e}^{-c\,(3\,t+2\,x)},\\
\\
C(x,t)=q_3\,\Big[a_1-\mathrm{e}^{\dfrac{3\,c\,(x+t)}{5}}\Big]^5\,\mathrm{e}^{-c\,(2\,t+3\,x)}.
  \end{array}
\right.
\end{equation}
It is observed from equations (\ref{uu4}), the line element (\ref{u21}) can be written in the following form:

\begin{equation}  \label{s12}
\begin{array}{ll}
ds_{113}^2=q_1^2\,\Big[a_1-\mathrm{e}^{\dfrac{3\,c\,(x+t)}{5}}\Big]^4\,\mathrm{e}^{-\dfrac{2\,c}{5}\,(5\,t+7\,x)}\,\Big(dx^2-dt^2\Big)
\\
\\
\,\,\,\,\,\,\,\,\,\,\,\,\,\,\,\,\,\,
+\Big[a_1-\mathrm{e}^{\dfrac{3\,c\,(x+t)}{5}}\Big]^{10}\,
\Big(q_2^2\,\mathrm{e}^{-2c\,(3\,t+2\,x)}\,dy^2+q_3^2\,\mathrm{e}^{-2\,c\,(2\,t+3\,x)}\,dz^2\Big),
\end{array}
\end{equation}
where $q_1$, $q_2$, $q_3$, $c$ and $a_1$ are an arbitrary constants.\\

\textbf{Solution (1.4):} We consider the solution correspondence the case (1.4). Without loss of generality, we can take, in this case, $d=-c$, $c_7=b_2\,a$, $c=2\,b_1$ and $n=\dfrac{1}{4}$. Now, by using (\ref{u42-15}),  (\ref{u42-8}), (\ref{u42-6}) and (\ref{u42-1}), we can find the solution as the following:
\begin{equation}\label{uu5}
\left\{
  \begin{array}{ll}
A(x,t)=q_1\,\exp\Big[\dfrac{1}{2\,b_2}\Big(2\,b_2^2\,x-b_1^2\,(x+t)+\mathrm{e}^{b_2\,(x+t)}\Big)\Big],\\
\\
B(x,t)=q_2\,\exp\Big[\dfrac{1}{b_2}\Big(b_1\big[2\,b_2\,x-(b_1+b_2)\,(x+t)\big]+\mathrm{e}^{b_2\,(x+t)}\Big)\Big],\\
\\
C(x,t)=q_3\,\exp\Big[\dfrac{1}{b_2}\Big(b_1\big[b_2\,(t-x)-b_1\,(x+t)\big]+\mathrm{e}^{b_2\,(x+t)}\Big)\Big].
  \end{array}
\right.
\end{equation}
It is observed from equations (\ref{uu5}), the line element (\ref{u21}) can be written in the following form:

\begin{equation}  \label{s14}
\begin{array}{ll}
ds_{14}^2=A^2\,\big(dx^2-dt^2\big)+B^2\,dy^2+C^2\,dz^2,
\end{array}
\end{equation}
where $q_1$, $q_2$, $q_3$, $b_1$ and $b_2$ are an arbitrary constants.\\

\textbf{Solution (1.5):} We consider the solution correspondence the case (1.5). Without loss of generality, we can take, in this case, $d=-c$, $c_7=0$, $c=2\,\sqrt{2}\,a_3$, and $n=\dfrac{1}{4}$. Now, by using (\ref{u42-15}),  (\ref{u42-8}), (\ref{u42-6}) and (\ref{u42-1}), we can find the solution as the following:
\begin{equation}\label{uu6}
\left\{
  \begin{array}{ll}
A(x,t)=q_1\,\exp\Big[\Big(\dfrac{x+t}{2}\Big)\,\Big(a_2+\sqrt{2}\,a_3+a_3^2\,(x+t)\Big)\Big],\\
\\
B(x,t)=q_2\,\exp\Big[a_2\,(x+t)+a_3^2\,(x+t)^2+2\,\sqrt{2}\,a_3\,x\Big],\\
\\
C(x,t)=q_3\,\exp\Big[a_2\,(x+t)+a_3^2\,(x+t)^2+2\,\sqrt{2}\,a_3\,t\Big].
  \end{array}
\right.
\end{equation}
It is observed from equations (\ref{uu6}), the line element (\ref{u21}) can be written in the following form:

\begin{equation}  \label{s15}
\begin{array}{ll}
ds_{15}^2=q_1^2\,\exp\Big[(x+t)\,\Big(a_2+\sqrt{2}\,a_3+a_3^2\,(x+t)\Big)\Big]\,\big(dx^2-dt^2\big)\\
+\exp\Big[2\,(x+t)\,\Big(a_2+a_3^2\,(x+t)\Big)\Big]\,\Big(q_2^2\,\mathrm{e}^{4\,\sqrt{2}\,a_3\,x}\,dy^2+q_3^2\,\mathrm{e}^{4\,\sqrt{2}\,a_3\,t}\,dz^2\Big),
\end{array}
\end{equation}
where $q_1$, $q_2$, $q_3$, $a_2$ and $a_3$ are an arbitrary constants.\\

\textbf{Case (2):} When $c_1\neq0$, the similarity variable and similarity functions can be written as the following:
\begin{equation}\label{u42-1-2}
\begin{array}{ll}
\xi=\dfrac{x+a}{t+b},\,\,\,\,\,B(x,t)=(x+a)^c\,\Psi(\xi),\,\,\,\,\,C(x,t)=(x+a)^d\,\Phi(\xi),
\end{array}
\end{equation}
where $a=\dfrac{c_2}{c_1}$, $b=\dfrac{c_3}{c_1}$, $c=\dfrac{c_4}{c_1}$ and $d=\dfrac{c_5}{c_1}$ are an arbitrary constants. Substituting the transformations (\ref{u42-1-2}) in the field Eqs. (\ref{u217})-(\ref{u218}) lead
to the following system of ordinary differential equations:
\begin{equation}\label{u43-1-2}
\begin{array}{ll}
\dfrac{\xi\Psi''+\big[1+c-c_9-2n\,(c+d)\big]\,\Psi'}{\xi\,\Psi}+\dfrac{\xi\Phi''+\big[1+c-c_9-2n\,(c+d)\big]\,\Phi'}{\xi\,\Phi}\\
\\
\,\,\,\,\,\,\,\,\,\,\,\,\,\,\,\,\,\,\,\,\,\,\,\,\,\,\,\,\,\,\,\,\,\,\,\,\,\,\,\,\,\,\,\,\,\,\,\,\,\,
\,\,\,\,\,\,\,\,\,\,\,\,\,\,\,\,\,\,\,\,\,\,\,\,\,\,\,\,\,\,\,\,\,\,\,\,\,\,\,\,\,\,\,\,\,\,\,\,\,\,
=2\,n\,\Big(\dfrac{\Psi'}{\Psi}+\dfrac{\Phi'}{\Phi}\Big)^2,
\end{array}
\end{equation}

\begin{equation}\label{u44-1-2}
\begin{array}{ll}
\dfrac{\Psi''}{\Psi}+\dfrac{\Phi''}{\Phi}+\dfrac{2\big[(2\,n+1)\,\xi^2-c\big]\,\Psi'}{(2\,n+1)\,(\xi^2-1)\,\xi\,\Psi}
+\dfrac{2\big[(2\,n+1)\,\xi^2-d\big]\,\Phi'}{(2\,n+1)\,(\xi^2-1)\,\xi\,\Phi}\\
\\
\,\,\,\,\,\,\,\,\,\,\,\,\,\,\,
-\dfrac{2\,n}{2\,n+1}\,\Big(\dfrac{\Psi'^2}{\Psi^2}+\dfrac{\Phi'^2}{\Phi^2}\Big)
=\dfrac{c\,(c-1)+d\,(d-1)-2\,c_9-2\,n\,(c+d)}{(2\,n+1)\,(\xi^2-1)\,\xi^2},
\end{array}
\end{equation}
where $f(x)=k\,(x+a)^{c_9}$, $k=c_8\,c_1^{c_9}$.\\

If one solves the system of second order non-linear ordinary differential equations (\ref{u43-1-2})-(\ref{u44-1-2}), he can obtain the exact solutions of the original Einstein field equations (\ref{u217})-(\ref{u218}) corresponding to reduction (\ref{u42-1-2}). The system (\ref{u43-1-2})-(\ref{u44-1-2}) is very difficult to solve in general form. This system may be solved in some special cases in the future work.\\

\section{Physical and geometrical properties of some models}

\textbf{For the Model (\ref{s111}):}

The expressions for energy density $\rho$, the string tension density $\lambda$, magnetic permeability $\bar{\mu}$ and The particle density $\rho_p$, for the model (\ref{s111}),  are given by:

\begin{equation}\label{uu1-1}
  \begin{array}{ll}
\rho(x,t)=-\lambda(x,t)=n_1\,\exp\Big[2\,c_7\Big(n\,(x+t)-x\Big)\Big]\,\sec^{n_1}\big[\xi\big]\,\Big(n_3\,\cos\big[\xi\big]-n_4\,\sin\big[\xi\big]\Big),
  \end{array}
\end{equation}

\begin{equation}\label{uu1-3}
  \begin{array}{ll}
\bar{\mu}(x,t)\,=\dfrac{\chi\,F^2_{12}(x)\,\exp\Big[c\,\Big(t-x-n_1\,(x+t)\Big)\Big]}{2\,c\,n_1\,q_2^2\,
\cos^{2\,n_1}\big[\xi\big]\,\Big(2\,K_1\,\tan\big[\xi\big]-c_7\Big)},
  \end{array}
\end{equation}

\begin{equation}\label{uu1-4}
  \begin{array}{ll}
\rho_p(x,t)\,=n_5\,\exp\Big[2\,c_7\Big(n\,(x+t)-x\Big)\Big]\,\cos^{n}\big[\xi\big],
  \end{array}
\end{equation}

where $n_3=\dfrac{c_7\,(c+c_7)+4\,K_1^2}{\chi\,q_1^2}$, $n_4=\dfrac{2\,c\,K_1}{\chi\,q_1^2}$, $n_5=\dfrac{2\,c^2}{q_1^2}$, $\xi=K_1\,\big(x+t\big)+\xi_0$ and $F_{12}(x)$ is an arbitrary function of the variable $x$.\\

The volume element is
\begin{equation}  \label{uu1-5}
V=q_1^2\,q_2\,q_3\,\exp\Big[\dfrac{c_7}{2}\Big(3\,x-t+3\,n_1\,(x+t)\Big)\Big]\,\cos^{3\,n_1-1}\big[\xi\big].
\end{equation}
The expansion scalar, which determines the volume behavior of the fluid, is given by:
\begin{equation}\label{uu1-6}
  \begin{array}{ll}
\Theta=\Big(\dfrac{5\,n_1-1}{4\,q_1}\Big)\,
\Big(c_7\,\cos\big[\xi\big]-2\,K_1\,\sin\big[\xi\big]\Big)\,\exp\Big[c_7\Big(n\,(x+t)-x\Big)\Big]\,\sec^{n\,n_1-1}\big[\xi\big],
  \end{array}
\end{equation}
The non-vanishing components of the shear tensor, $\sigma_i^j$, are:
\begin{equation}\label{uu1-7}
  \begin{array}{ll}
\sigma_1^1\,=\dfrac{2\,(n_1+1)\,\Theta}{3\,(1-5\,n_1)}.
  \end{array}
\end{equation}

\begin{equation}\label{uu1-8}
  \begin{array}{ll}
\sigma_2^2\,=\dfrac{\Theta}{3\,(5\,n_1-1)}\Big[n_1+1-\dfrac{6\,c\,\cos\big[\xi\big]}{c_7\,\cos\big[\xi\big]-2\,K_1\,\sin\big[\xi\big]}\Big],
  \end{array}
\end{equation}

\begin{equation}\label{uu1-9}
  \begin{array}{ll}
\sigma_3^3\,=-\Big(\sigma_1^1+\sigma_2^2\Big).
  \end{array}
\end{equation}

The shear scalar is:
\begin{equation}\label{uu1-10}
  \begin{array}{ll}
\sigma^2\,=\dfrac{\Big[n_6+n_7\,\cos\big[\xi\big]-4\,c_7\,K_1\,\sin\big[\xi\big]\Big]\,\Theta^2}{6\,(1-5\,n_1)^2\,
\Big(c_7\,\cos\big[\xi\big]-2\,K_1\,\sin\big[\xi\big]\Big)^2},
  \end{array}
\end{equation}
where $n_6=(n_1^2-10\,n_1+1)(c_7^2+4\,K_1^2)$ and $n_7=(n_1^2+1)(c_7^2-4\,K_1^2)-2\,n_1\,(5\,c_7^2+28\,K_1^2)$.\\

The acceleration vector is given by:
\begin{equation}\label{uu1-11}
  \begin{array}{ll}
\dot{u}_i=\dfrac{1}{4}\Big(c_7\,(n_1+3)+2\,K_1\,(1-n_1)\,\tan\big[\xi\big],0,0,0\Big).
  \end{array}
\end{equation}
The deceleration parameter is given by \cite{dec1,rayc1}
\begin{equation}\label{uu1-12}
  \begin{array}{ll}
\mathbf{q}&=-3\,\Theta^2\,\Big(\Theta_{;i}\,u^{i}+\dfrac{1}{3}\,\Theta^2\Big)\\
&=\dfrac{(5\,n_1-1)^3}{256\,q_1^4}\exp\Big[4\,c_7\Big(n\,(x+t)-x\Big)\Big]\,\Big(c_7\,\cos\big[\xi\big]-2\,K_1\,\sin\big[\xi\big]\Big)
\\
&\,\,\,\,\,\times\Bigg(c^2-c_7^2+20\,K_1^1-(n_1+1)\Big[(c_7^2-4\,K_1^2)\,\cos\big[\xi\big]-4\,c_7\,K_1\,\sin\big[\xi\big]\Big]\Bigg).
  \end{array}
\end{equation}

\textbf{Remark (3):} It is worth noting that: If we put the following transformation
$$
K_1\,\rightarrow\,\imath\,K_2,\,\,\,\sin\,\rightarrow\,\sinh,\,\,\,\cos\,\rightarrow\,\cosh,
$$
in the model (\ref{s111}), we have obtained the model (\ref{s112}), where $\imath=\sqrt{-1}$. Therefore, we can find the physical properties of the model correspondence to case (1.1.2) by putting the above transformation in the the model correspondence to case (1.1.1).\\

\textbf{For the Model (\ref{s113}):}

The expressions for energy density $\rho$, the string tension density $\lambda$, magnetic permeability $\bar{\mu}$ and The particle density $\rho_p$, for the model (\ref{s113}),  are given by:

\begin{equation}\label{uu3-1}
  \begin{array}{ll}
\rho(x,t)=-\lambda(x,t)=\dfrac{4\,d_1}{\chi\,q_1^2}\,(x+t)^{d_2^2}\,\Big[d_2^2-n_+\,(x+t)\Big]\,\exp\Big[-4\,d_1\,d_2\,(n\,t+n_1\,x)\Big],
  \end{array}
\end{equation}

\begin{equation}\label{uu3-3}
  \begin{array}{ll}
\bar{\mu}(x,t)\,=\dfrac{\chi\,F^2_{12}(x)\,(x+t)^{1+2\,d_2^2}\,\,\exp\Big[-4\,(n_+\,t+n_-\,x)\Big]}{8\,d_1\,d_2\,q_2^2\,\Big[d_1\,(x+t)-d_2\Big]},
  \end{array}
\end{equation}

\begin{equation}\label{uu3-4}
  \begin{array}{ll}
\rho_p(x,t)\,=\dfrac{8\,d_1^2}{q_1^2}\,(x+t)^{1+d_2^2}\,\exp\Big[-4\,d_1\,d_2\,(n\,t+n_1\,x)\Big],
  \end{array}
\end{equation}

where $F_{12}(x)$ is an arbitrary function of the variable $x$.\\

The volume element is
\begin{equation}  \label{uu3-5}
V=q_1^2\,q_2\,q_3\,(x+t)^{-1-3\,d_2^2}\,\exp\Big[d_1\,\Big((1+3\,d_2^2)\,t+3\,(d_2^2-1)\,x\Big)\Big].
\end{equation}
The expansion scalar, which determines the volume behavior of the fluid, is given by:
\begin{equation}\label{uu3-6}
  \begin{array}{ll}
\Theta=\dfrac{(5\,d_2^2+1)\,\Big[d_1\,(x+t)-d_2\Big]}{2\,d_2\,q_1}\,(x+t)^{\dfrac{d_2^2-1}{2}}\,\exp\Big[-2\,d_1\,d_2\,(n\,t+n_1\,x)\Big],
  \end{array}
\end{equation}
The non-vanishing components of the shear tensor, $\sigma_i^j$, are:
\begin{equation}\label{uu3-7}
  \begin{array}{ll}
\sigma_1^1\,=\dfrac{2\,(1-d_2^2)\,\Theta}{3\,(1-5\,d_2^2)}.
  \end{array}
\end{equation}

\begin{equation}\label{uu3-8}
  \begin{array}{ll}
\sigma_2^2\,=\dfrac{\Big[d_2\,(1+d_2)+d_1\,(1-6\,d_2-d_2^2)\,(x+t)\Big]\,\Theta}{3\,(1+5\,d_2^2)\,\Big[d_2-d_1\,(x+t)\Big]},
  \end{array}
\end{equation}

\begin{equation}\label{uu3-9}
  \begin{array}{ll}
\sigma_3^3\,=-\Big(\sigma_1^1+\sigma_2^2\Big).
  \end{array}
\end{equation}

The shear scalar is:
\begin{equation}\label{uu3-10}
  \begin{array}{ll}
\sigma^2\,=\dfrac{\Theta^2}{3\,(1+5\,d_2^2)^2}\,\Bigg[1+d_2^2\,(10+d_2^2)+\dfrac{12\,d_2^3\Big(2\,d_1\,(x+t)-2\,d_2+1\Big)}{\Big(d_1\,(x+t)-\,d_2\Big)^2}\Bigg].
  \end{array}
\end{equation}

The acceleration vector is given by:
\begin{equation}\label{uu3-11}
  \begin{array}{ll}
\dot{u}_i=2\Big(n_1\,d_1-\dfrac{n\,d_2}{x+t},0,0,0\Big).
  \end{array}
\end{equation}
The deceleration parameter is given by:
\begin{equation}\label{uu3-12}
  \begin{array}{ll}
\mathbf{q}&=-\dfrac{(5\,d_2^2+1)^3}{8\,d_2^4\,q_1^4}\,(x+t)^{2\,(d_2^2-1)}\,\exp\Big[-8\,d_1\,d_2\,(n\,t+n_1\,x)\Big]\,
\\
&\,\,\,\,\,\,\,\,\,\,\,\,\,\,\,\times\Bigg(d_2^2\,(d_2^2+2)+d_1\,(d_2^2-1)\,(x+t)\,\Big[d_1\,(x+t)-2\,d_2\Big]\Bigg).
  \end{array}
\end{equation}

\textbf{For the Model (\ref{s12}):}

The expressions for energy density $\rho$, the string tension density $\lambda$, magnetic permeability $\bar{\mu}$ and the particle density $\rho_p$, for the model (\ref{s12}),  are given by:

\begin{equation}\label{uu4-1}
  \begin{array}{ll}
\rho(x,t)\,=\,-\lambda(x,t)=\dfrac{6\,a_1\,c^2\,\mathrm{e}^{\dfrac{2\,c\,(5\,t+7\,x)}{5}}}{\chi\,q_1}\,\Big[a_1-\mathrm{e}^{\dfrac{3\,c\,(x+t)}{5}}\Big]^{-5},
  \end{array}
\end{equation}

\begin{equation}\label{uu4-3}
  \begin{array}{ll}
\bar{\mu}(x,t)\,=\,\dfrac{\chi\,F^2_{12}(x)\,\mathrm{e}^{2\,c\,(3\,t+2\,x)}}{2\,c^2\,q_2^2}\,\Big[a_1-\mathrm{e}^{\dfrac{3\,c\,(x+t)}{5}}\Big]^{-9}\,
\Big[5\,a_1-\mathrm{e}^{\dfrac{3\,c\,(x+t)}{5}}\Big]^{-1},
  \end{array}
\end{equation}

\begin{equation}\label{uu4-4}
  \begin{array}{ll}
\rho_p(x,t)\,=\,\dfrac{2\,c\,\mathrm{e}^{\dfrac{2\,c\,(5\,t+7\,x)}{5}}}{q_1^2}\,\Big[a_1-\mathrm{e}^{\dfrac{3\,c\,(x+t)}{5}}\Big]^{-4},
  \end{array}
\end{equation}

where $F_{12}(x)$ is an arbitrary function of the variable $x$.\\

The volume element is
\begin{equation}  \label{uu4-5}
V=q_1^2\,q_2\,q_3\,\,\mathrm{e}^{\dfrac{-\,c\,(35\,t+39\,x)}{5}}\,\Big[a_1-\mathrm{e}^{\dfrac{3\,c\,(x+t)}{5}}\Big]^{14}.
\end{equation}
The expansion scalar, which determines the volume behavior of the fluid, is given by:
\begin{equation}\label{uu4-6}
  \begin{array}{ll}
\Theta=-\dfrac{6\,c}{5\,q_1}\,\mathrm{e}^{\dfrac{c\,(5\,t+7\,x)}{5}}\,\Big[5\,a_1+\mathrm{e}^{\dfrac{3\,c\,(x+t)}{5}}\Big]\,
\Big[a_1-\mathrm{e}^{\dfrac{3\,c\,(x+t)}{5}}\Big]^{-3},
  \end{array}
\end{equation}
The non-vanishing components of the shear tensor, $\sigma_i^j$, are:
\begin{equation}\label{uu4-7}
  \begin{array}{ll}
\sigma_1^1\,=\dfrac{\Theta}{6},
  \end{array}
\end{equation}

\begin{equation}\label{uu4-8}
  \begin{array}{ll}
\sigma_2^2\,=\dfrac{\Theta}{6}\,\Big[5\,a_1-2\,\mathrm{e}^{\dfrac{3\,c\,(x+t)}{5}}\Big]\,
\Big[5\,a_1+\mathrm{e}^{\dfrac{3\,c\,(x+t)}{5}}\Big]^{-1},
  \end{array}
\end{equation}

\begin{equation}\label{uu4-9}
  \begin{array}{ll}
\sigma_3^3\,=-\Big(\sigma_1^1+\sigma_2^2\Big).
  \end{array}
\end{equation}

The shear scalar is:
\begin{equation}\label{uu4-10}
  \begin{array}{ll}
\sigma^2\,=\dfrac{\Theta^2}{36}\,\Big[25\,a_1^2-5\,a_1\,\mathrm{e}^{\dfrac{3\,c\,(x+t)}{5}}+7\,\mathrm{e}^{\dfrac{6\,c\,(x+t)}{5}}\Big]\,
\Big[5\,a_1+\mathrm{e}^{\dfrac{3\,c\,(x+t)}{5}}\Big]^{-2}.
  \end{array}
\end{equation}

The acceleration vector is given by:
\begin{equation}\label{uu4-11}
  \begin{array}{ll}
\dot{u}_i=-\dfrac{c}{7}\Big[7\,a_1-\mathrm{e}^{\dfrac{3\,c\,(x+t)}{5}}\Big]\,\Big[a_1-\mathrm{e}^{\dfrac{3\,c\,(x+t)}{5}}\Big]^{-1}\,
\Big(1,0,0,0\Big).
  \end{array}
\end{equation}
The deceleration parameter is given by:
\begin{equation}\label{uu4-12}
  \begin{array}{ll}
\mathbf{q}&=-\dfrac{18\,c^2\,\Theta^2}{25\,q_1^2}\,\mathrm{e}^{\dfrac{2\,c\,(5\,t+7\,t)}{5}}\,
\Big[a_1-\mathrm{e}^{\dfrac{3\,c\,(x+t)}{5}}\Big]^{-6}\\
&\,\,\,\,\,\,\,\,\,\,\,\,\,\,\,\,\,\,\,\,\times\Big[25\,a_1^2-8\,a_1\,\mathrm{e}^{\dfrac{3\,c\,(x+t)}{5}}+\,\mathrm{e}^{\dfrac{6\,c\,(x+t)}{5}}\Big].
  \end{array}
\end{equation}

\textbf{For the Model (\ref{s14}):}

The expressions for energy density $\rho$, the string tension density $\lambda$, magnetic permeability $\bar{\mu}$ and the particle density $\rho_p$, for the model (\ref{s14}),  are given by:

\begin{equation}\label{uu5-1}
  \begin{array}{ll}
\rho(x,t)\,=\,-\lambda(x,t)=-\dfrac{4\,b_1}{\chi\,b_2\,q_1^2}\,\Big[b_1\,(b_1+b_2)-b_2\,\mathrm{e}^{b_2\,(x+t)}\Big]\\
\\
\,\,\,\,\,\,\,\,\,\,\,\,\,\,\,\,\,\,\,\,\,\,\,\,\,\,\,\,\,\,\,\,\,\,\,\,\,\,\,\,\,\,
\times\exp\Big[\dfrac{1}{2\,b_2}\,\Big(b_1^2\,(x+t)-2\,b_2^2-\mathrm{e}^{b_2\,(x+t)}\Big)\Big],
  \end{array}
\end{equation}

\begin{equation}\label{uu5-3}
  \begin{array}{ll}
\bar{\mu}(x,t)=\dfrac{\chi\,b_2\,F^2_{12}(x)}{8\,b_1\,q_2^2\,\Big[b_1-b_2\,\mathrm{e}^{b_2\,(x+t)}\Big]}\,
\exp\Big[\dfrac{2}{b_2}\,\Big[b_1\Big(b_2\,(t-x)+b_1\,(x+t)\Big)-\mathrm{e}^{b_2\,(x+t)}\Big]\Big],
  \end{array}
\end{equation}

\begin{equation}\label{uu5-4}
  \begin{array}{ll}
\rho_p(x,t)\,=\,\dfrac{8\,b_1^2}{q_1^2}\,\exp\Big[\dfrac{1}{b_2}\,\Big(b_1^2(x-t)-2\,b_2^2\,x-\mathrm{e}^{b_2\,(x+t)}\Big)\Big],
  \end{array}
\end{equation}

where $F_{12}(x)$ is an arbitrary function of the variable $x$.\\

The volume element is
\begin{equation}  \label{uu5-5}
V=q_1^2\,q_2\,q_3\,\exp\Big[\dfrac{1}{b_2}\,\Big(2\,b_2^2\,x-3\,b_1^2(x-t)+3\,\mathrm{e}^{b_2\,(x+t)}\Big)\Big].
\end{equation}
The expansion scalar, which determines the volume behavior of the fluid, is given by:
\begin{equation}\label{uu5-6}
  \begin{array}{ll}
\Theta=\,\dfrac{5}{2\,b_2\,q_1}\,\Big[b_2\,\mathrm{e}^{b_2\,(x+t)}-b_1^2\Big]\,
\exp\Big[\dfrac{1}{2\,b_2}\,\Big(b_1^2\,(x+t)-2\,b_2^2\,x-\mathrm{e}^{b_2\,(x+t)}\Big)\Big],,
  \end{array}
\end{equation}
The non-vanishing components of the shear tensor, $\sigma_i^j$, are:
\begin{equation}\label{uu5-7}
  \begin{array}{ll}
\sigma_1^1\,=-\dfrac{2\,\Theta}{15},
  \end{array}
\end{equation}

\begin{equation}\label{uu5-8}
  \begin{array}{ll}
\sigma_2^2\,=\dfrac{\Theta}{15}\,\Big(\dfrac{b_1^2+6\,b_1\,b_2-b_2\,\mathrm{e}^{b_2\,(x+t)}}{b_1^2-b_2\,\mathrm{e}^{b_2\,(x+t)}}\Big),
  \end{array}
\end{equation}

\begin{equation}\label{uu5-9}
  \begin{array}{ll}
\sigma_3^3\,=-\Big(\sigma_1^1+\sigma_2^2\Big).
  \end{array}
\end{equation}

The shear scalar is:
\begin{equation}\label{uu5-10}
  \begin{array}{ll}
\sigma^2\,=\dfrac{\Theta^2}{25}\,\Big[\dfrac{1}{3}+4\,b_1^2\,b_2^2\,\Big(b_1^2-b_2\,\mathrm{e}^{b_2\,(x+t)}\Big)^{-2}\Big].
  \end{array}
\end{equation}

The acceleration vector is given by:
\begin{equation}\label{uu5-11}
  \begin{array}{ll}
\dot{u}_i=\dfrac{1}{2\,d_1}\Big[2\,b_2^2-b_1^2+b_2\,\mathrm{e}^{b_2\,(x+t)}\Big]\,
\Big(1,0,0,0\Big).
  \end{array}
\end{equation}
The deceleration parameter is given by:
\begin{equation}\label{uu4-12}
  \begin{array}{ll}
\mathbf{q}=-\dfrac{5\,\Theta^2}{2\,b_2^2\,q_1^2}\,\Big[b_1^2+b_2\,(3\,b_2^2-2\,b_1^2)\,\mathrm{e}^{b_2^2\,(x+t)}+2\,b_2^2\,\mathrm{e}^{b_2\,(x+t)}\Big]\\
\\
\,\,\,\,\,\,\,\,\,\,\,\,\,\,\,\,\,\,\,\,\,\,\,\,\,\,\,\,\,\,\,\,
\times\exp\Big[-\dfrac{1}{b_2}\,\Big(2\,b_2^2-b_1^2\,(x+t)+\mathrm{e}^{b_2\,(x+t)}\Big)\Big].
  \end{array}
\end{equation}

\textbf{For the Model (\ref{s15}):}

The expressions for energy density $\rho$, the string tension density $\lambda$, magnetic permeability $\bar{\mu}$ and the particle density $\rho_p$, for the model (\ref{s15}),  are given by:

\begin{equation}\label{uu6-1}
  \begin{array}{ll}
\rho(x,t)\,=\,-\lambda(x,t)=\dfrac{4\,\sqrt{2}\,a_3}{\chi\,q_1^2}\,\Big[a_2+2\,a_3^2\,(x+t)\Big]\,
\exp\Big[-(x+t)\Big(a_2+\sqrt{2}\,a_3+a_3^2\,(x+t)\Big)\Big],
  \end{array}
\end{equation}

\begin{equation}\label{uu6-3}
  \begin{array}{ll}
\bar{\mu}(x,t)\,=\,-\dfrac{\chi\,F^2_{12}(x)}{8\,\sqrt{2}\,a_3\,q_2^2}\,\Big[\sqrt{2}\,a_3+a_2+2\,a_2^2\,(x+t)\Big]^{-1}\\
\\
\,\,\,\,\,\,\,\,\,\,
\times\exp\Big[-2\Big(2\,\sqrt{2}\,a_3+a_2\,(x+t)+a_3^2\,(x+t)^2\Big)\Big],
  \end{array}
\end{equation}

\begin{equation}\label{uu6-4}
  \begin{array}{ll}
\rho_p(x,t)\,=\,\dfrac{16\,a_3^2}{q_1^2}\,\exp\Big[-(x+t)\,\Big(\sqrt{2}\,a_3+a_2+a_3^2\,(x+t)\Big)\Big],
  \end{array}
\end{equation}

where $F_{12}(x)$ is an arbitrary function of the variable $x$.\\

The volume element is
\begin{equation}  \label{uu6-5}
V=q_1^2\,q_2\,q_3\,\exp\Big[3(x+t)\,\Big(\sqrt{2}\,a_3+a_2+a_3^2\,(x+t)\Big)\Big].
\end{equation}
The expansion scalar, which determines the volume behavior of the fluid, is given by:
\begin{equation}\label{uu6-6}
  \begin{array}{ll}
\Theta=\,\dfrac{5}{2\,q_1}\,\Big[\sqrt{2}\,a_3+a_2+2\,a_3^2\,(x+t)\Big]\,
\exp\Big[-\Big(\dfrac{x+t}{2}\Big)\,\Big(\sqrt{2}\,a_3+a_2+a_3^2\,(x+t)\Big)\Big],
  \end{array}
\end{equation}
The non-vanishing components of the shear tensor, $\sigma_i^j$, are:
\begin{equation}\label{uu6-7}
  \begin{array}{ll}
\sigma_1^1\,=-\dfrac{2\,\Theta}{15},
  \end{array}
\end{equation}

\begin{equation}\label{uu6-8}
  \begin{array}{ll}
\sigma_2^2\,=\dfrac{\Theta}{15}\,\Big(\dfrac{a_2-5\,\sqrt{2}\,a_3+2\,a_3^2\,(x+t)}{a_2+\sqrt{2}\,a_3+2\,a_3^2\,(x+t)}\Big),
  \end{array}
\end{equation}

\begin{equation}\label{uu6-9}
  \begin{array}{ll}
\sigma_3^3\,=-\Big(\sigma_1^1+\sigma_2^2\Big).
  \end{array}
\end{equation}

The shear scalar is:
\begin{equation}\label{uu6-10}
  \begin{array}{ll}
\sigma^2\,=\dfrac{\Theta^2}{25}\,\Big[\dfrac{1}{3}+8\,a_3^2\,\Big(a_2+\sqrt{2}\,a_3+2\,a_3^2\,(x+t)\Big)^{-2}\Big].
  \end{array}
\end{equation}

The acceleration vector is given by:
\begin{equation}\label{uu6-11}
  \begin{array}{ll}
\dot{u}_i=\Big[\dfrac{a_3}{\sqrt{2}}+\dfrac{a_2}{2}+a_3^2\,(x+t)\Big]\,
\Big(1,0,0,0\Big).
  \end{array}
\end{equation}
The deceleration parameter is given by:
\begin{equation}\label{uu6-12}
  \begin{array}{ll}
\mathbf{q}=-\dfrac{5\,\Theta^2}{2\,q_1^2}\,\Big[2\,\sqrt{2}\,a_3\,a_2+a_2^2+8\,a_3^2+4\,a_3^2\,(\sqrt{2}\,a_3+a_2)\,(x+t)+4\,a_3^4\,(x+t)^2\Big]\\
\\
\,\,\,\,\,\,\,\,\,\,\,\,\,\,\,\,\,\,\,\,\,\,\,\,\,\,\,\,\,\,\,\,
\times\exp\Big[-(x+t)\Big(a_2+\sqrt{2}\,a_3+a_3^2\,(x+t)\Big)\Big].
  \end{array}
\end{equation}

\section {Conclusion}

In the paper, we have  derived some new invariant solutions of
Einstein-Maxwell's field equations for string fluid as source of
matter in cylindrically symmetric space-time  with Variable Magnetic
Permeability. Different set of solutions  are found using different
values of the parameters. Note that  the cosmological solutions are
physically viable  for the following reasons:

  [i] energy density is positive and decreasing with the increase of the time

   [ii]  volume of the Universe is increasing due the expanding
   nature of the Universe

[iii] deceleration parameter should be negative as recent
observations indicate that our Universe is accelerating

[iv] $\frac{\sigma}{\theta}$  will be vanished at large time as  the
Universe may got
 isotropized in some later time

[v] solutions must be non singular as existence of Big-bang
singularity is one of the basic failures of general theory of
relativity

The models (61), (65), (67) and (69)  do not meet the above
criterion (i). Here, in the  models (61) and  (65),   densities
increase  with time  whereas in models  (67)   particle density is
increasing and  in model (69) particle density is negative (Fig-1).
Therefore, these models are not physically interesting. On the other
hand, the model (71) is very much acceptable   as it describes more
or less observable Universe. It is to be noted that our procedure of
solving the field equations (symmetry analysis method) is completely
different compare to usual methods available in literature. The
derived model starts expanding without Big Bang singularity (
fig-2). Also, from the theoretical perspective, the present model
can be a viable model to explain the  acceleration of the Universe.
In other words, the solution presented here can be one of the
potential candidates   to describe the observed Universe. In our
model,  it seems magnetic field with negative  magnetic Permeability
(Fig-3) is responsible to provide   accelerated  as well as
singularity free Universe. Nowadays, negative magnetic permeability
is not an  impossible event rather it can be found in split ring
resonator (SRR)   in the visible light region \cite{atsu1}. We also
note that $\frac{\sigma}{\theta}$  will be vanished at large time
(Fig-2). This means     the Universe may got
 isotropized in some later time. Again, one can  assume  that negative  magnetic Permeability is responsible for
the isotropisation.  A detail discussions of all basic cosmological
 constraints is beyond the scope   in the present
paper. We hope it will be taken care in  a future project.

\begin{figure*}[thbp]
\begin{tabular}{rl}
\includegraphics[width=4cm]{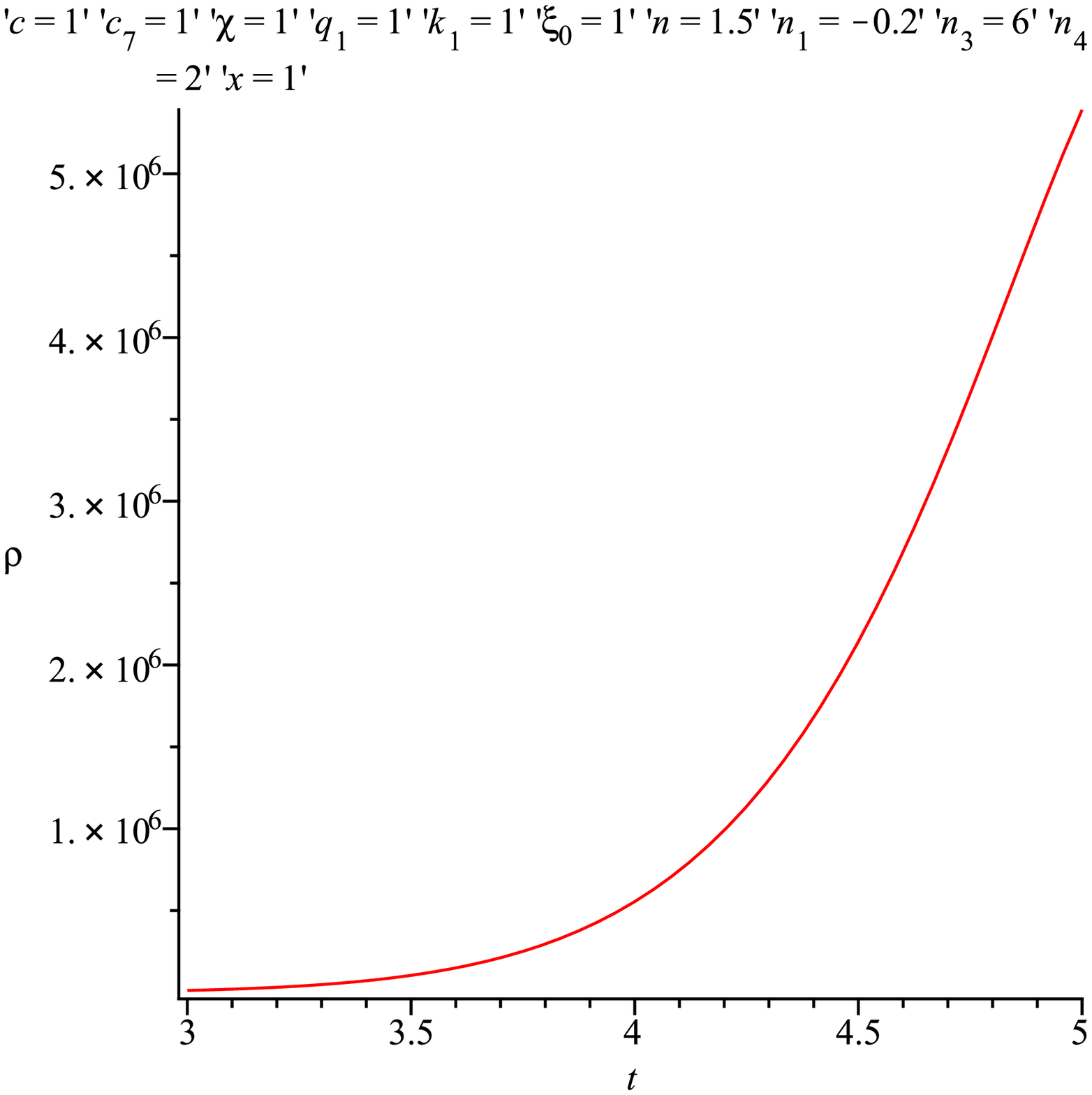}&
\includegraphics[width=4cm]{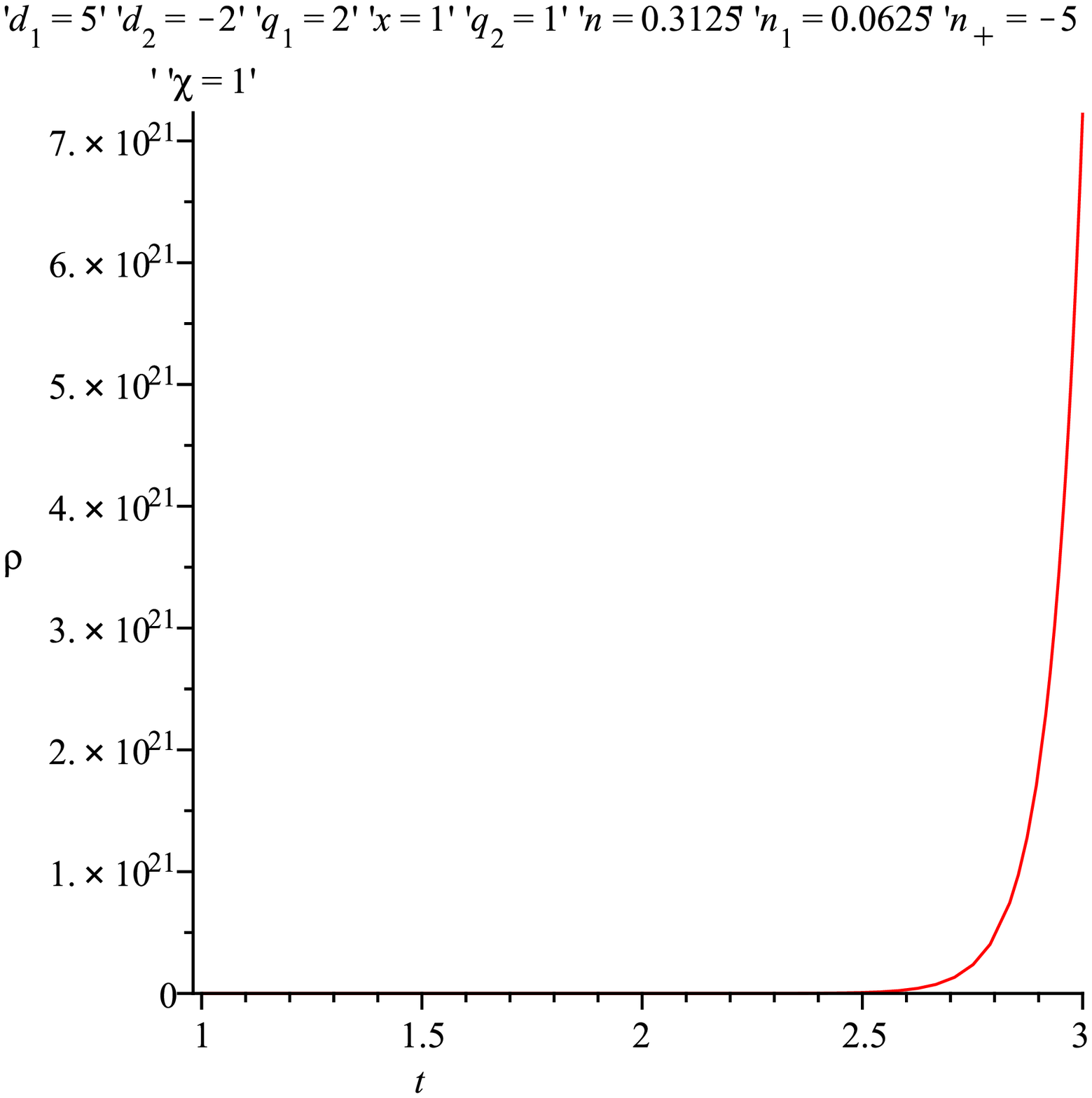}
\includegraphics[width=4cm]{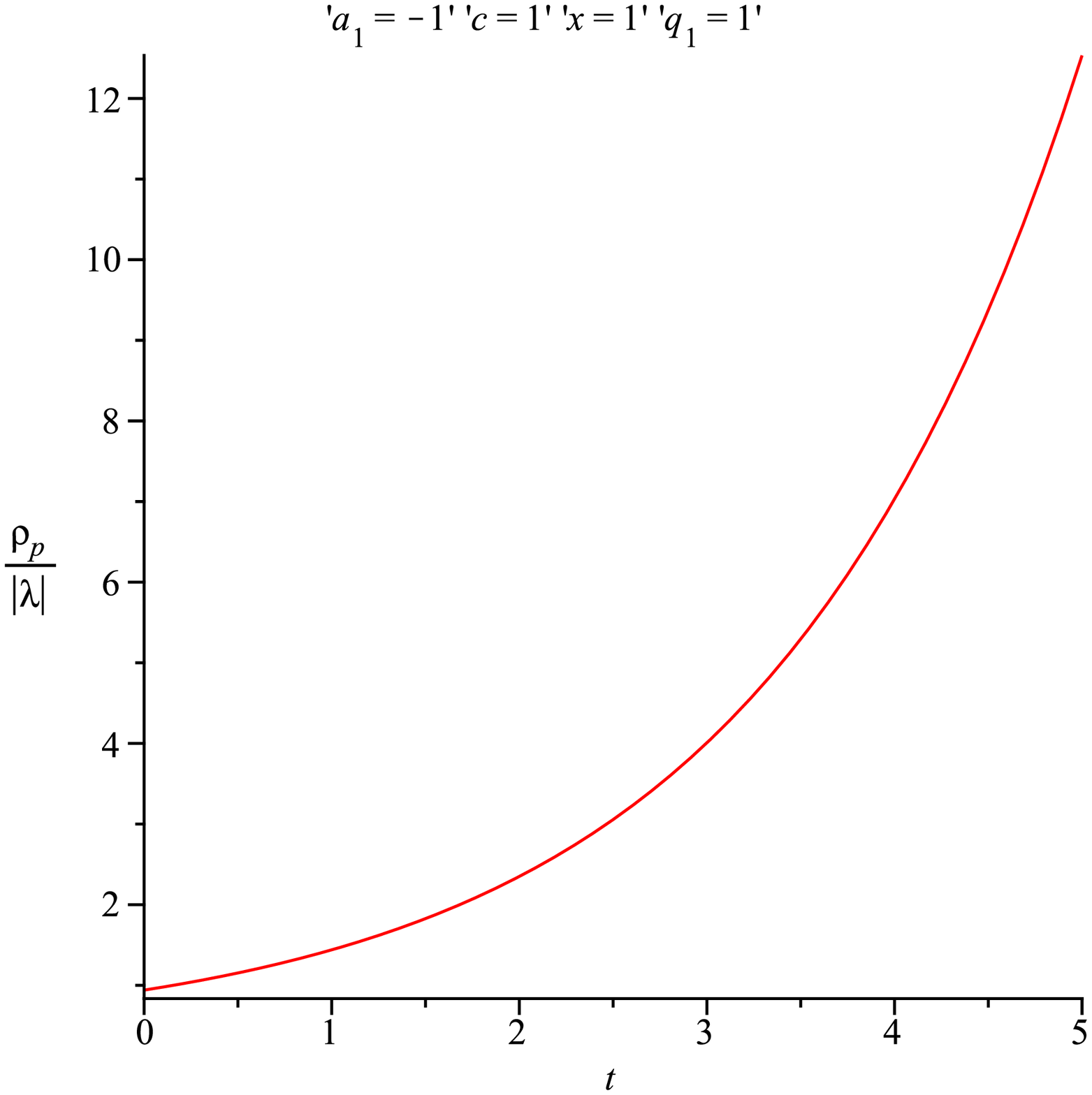}
\includegraphics[width=4cm]{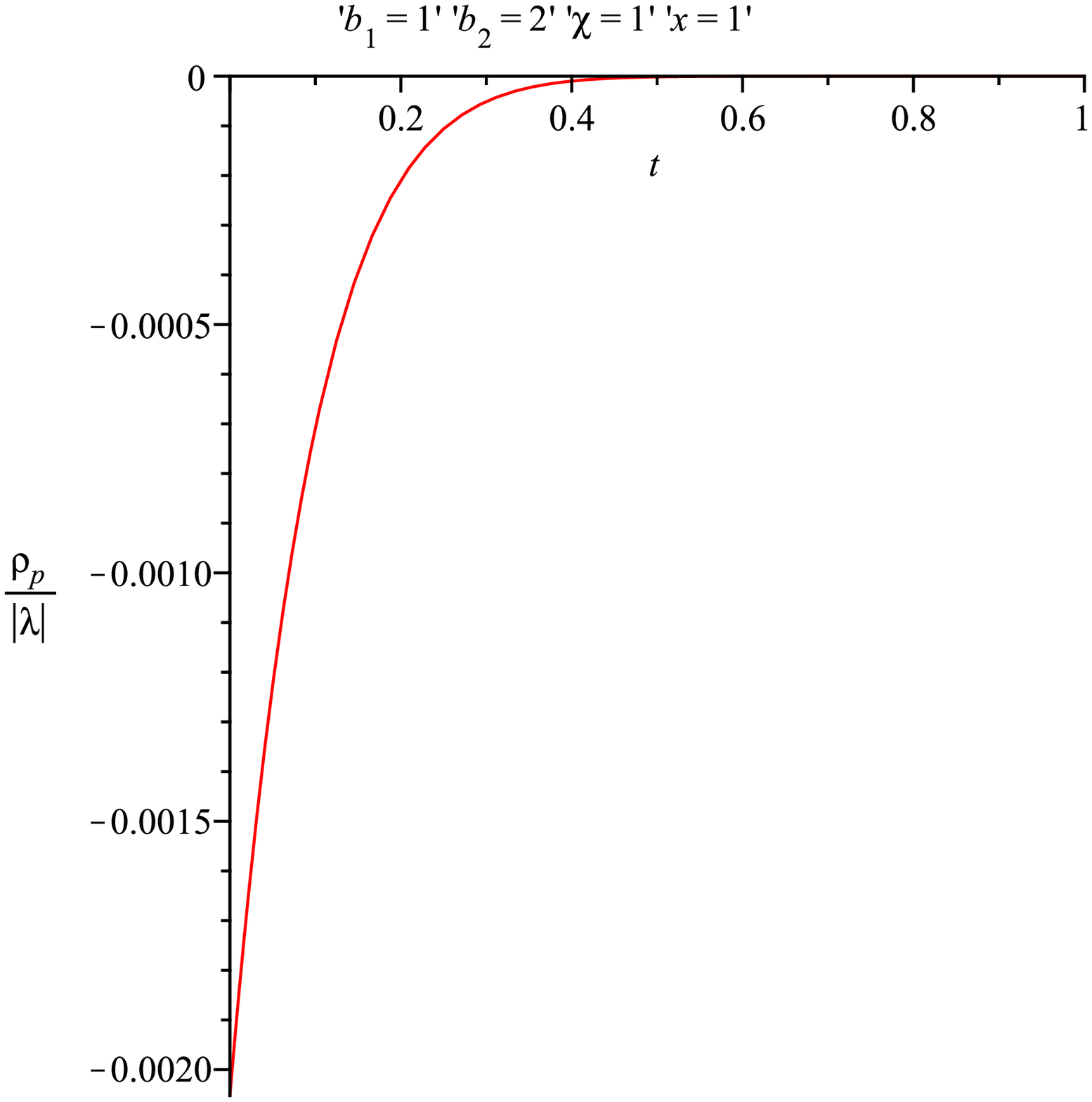}
 \\
\end{tabular}
\caption{ (\textit{Left})  The variation of energy density with
respect to time for the model(61). (\textit{First Middle})The
variation of energy density with respect to time for the model(65).
 (\textit{Second Middle}) The variation of particle density
with respect to time for model (67). (\textit{Right}) The variation
of particle density with respect to time for model (69). }
\end{figure*}
\begin{figure*}[thbp]
\begin{tabular}{rl}
\includegraphics[width=4.5cm]{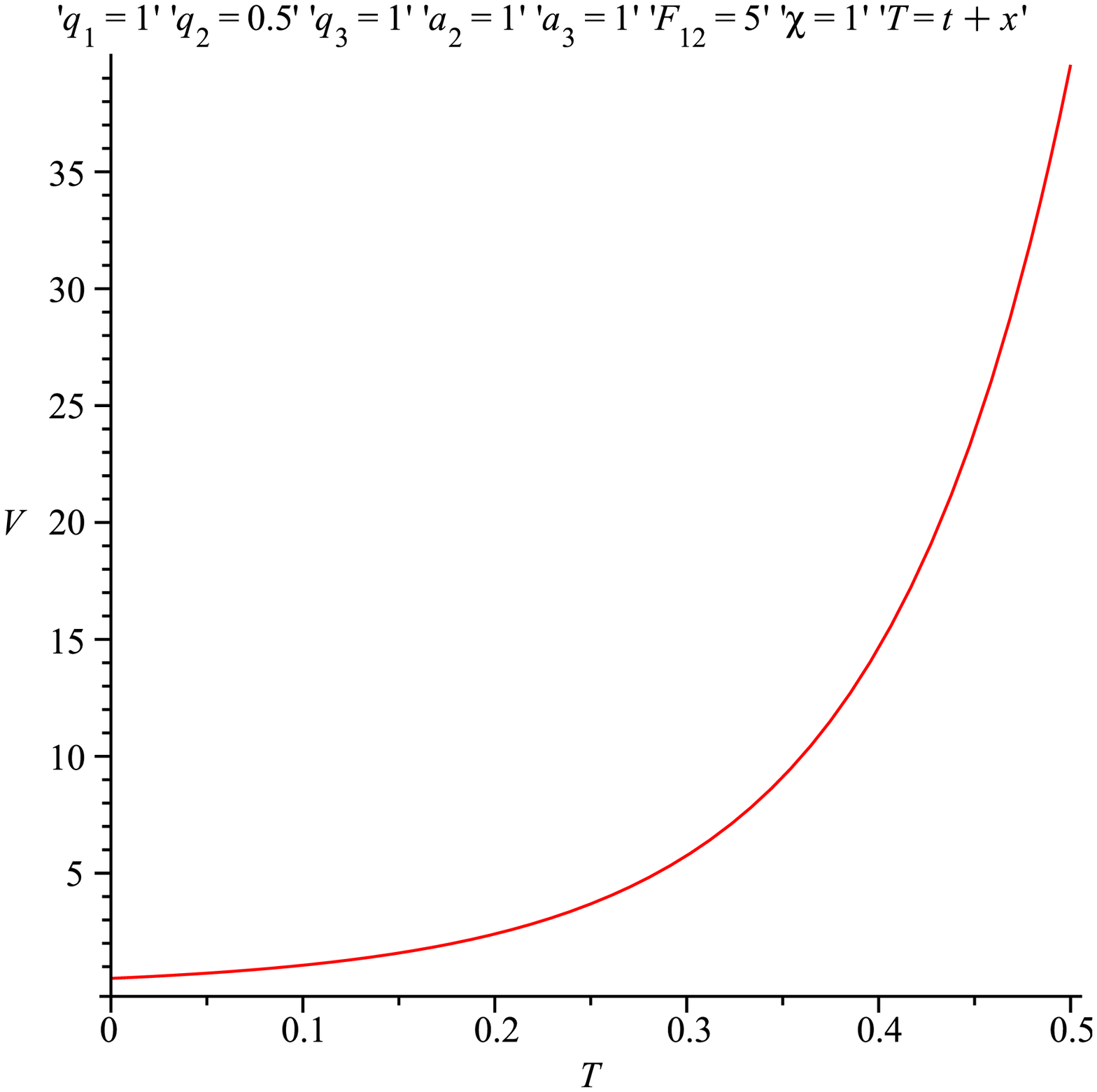}&
\includegraphics[width=4.5cm]{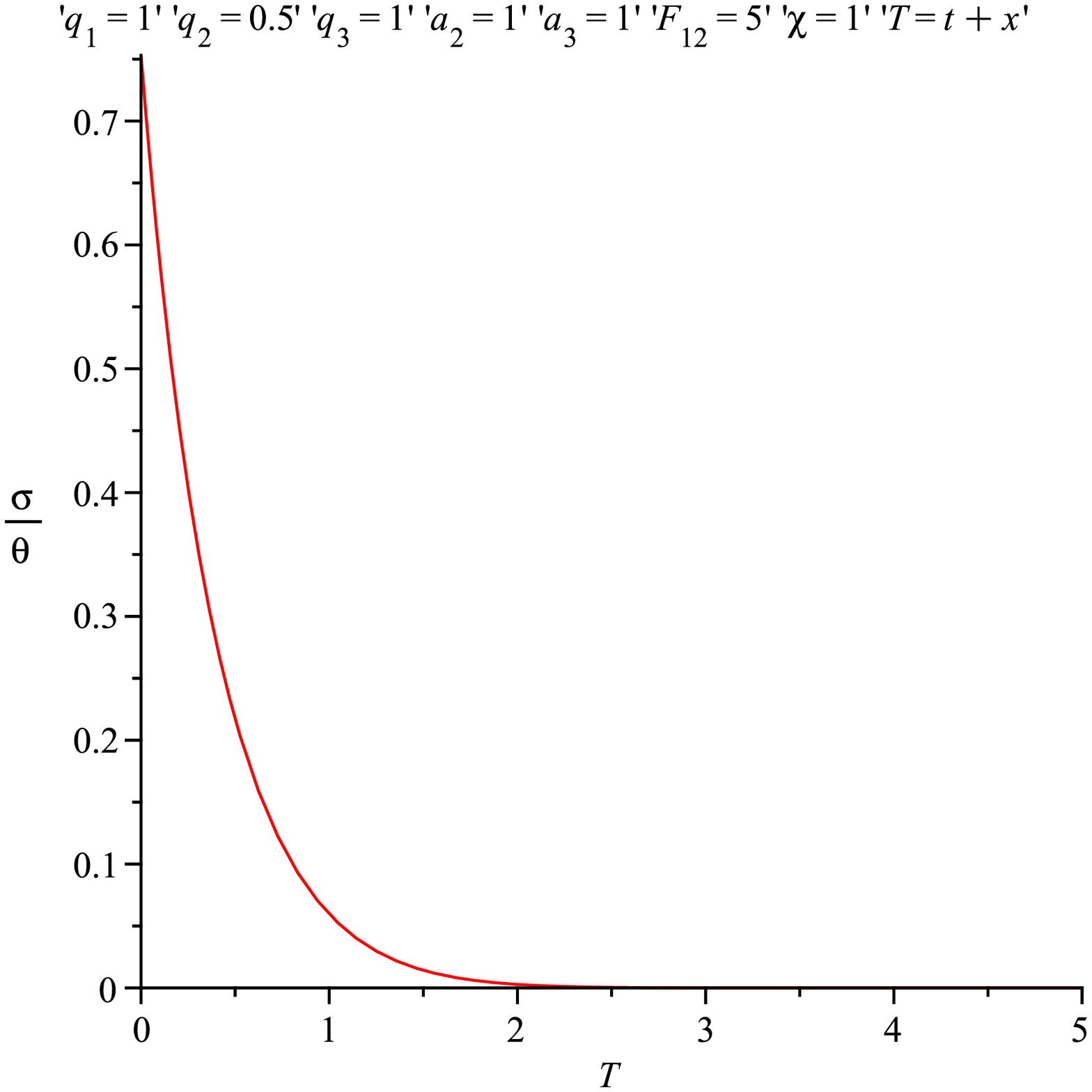}
\includegraphics[width=4.5cm]{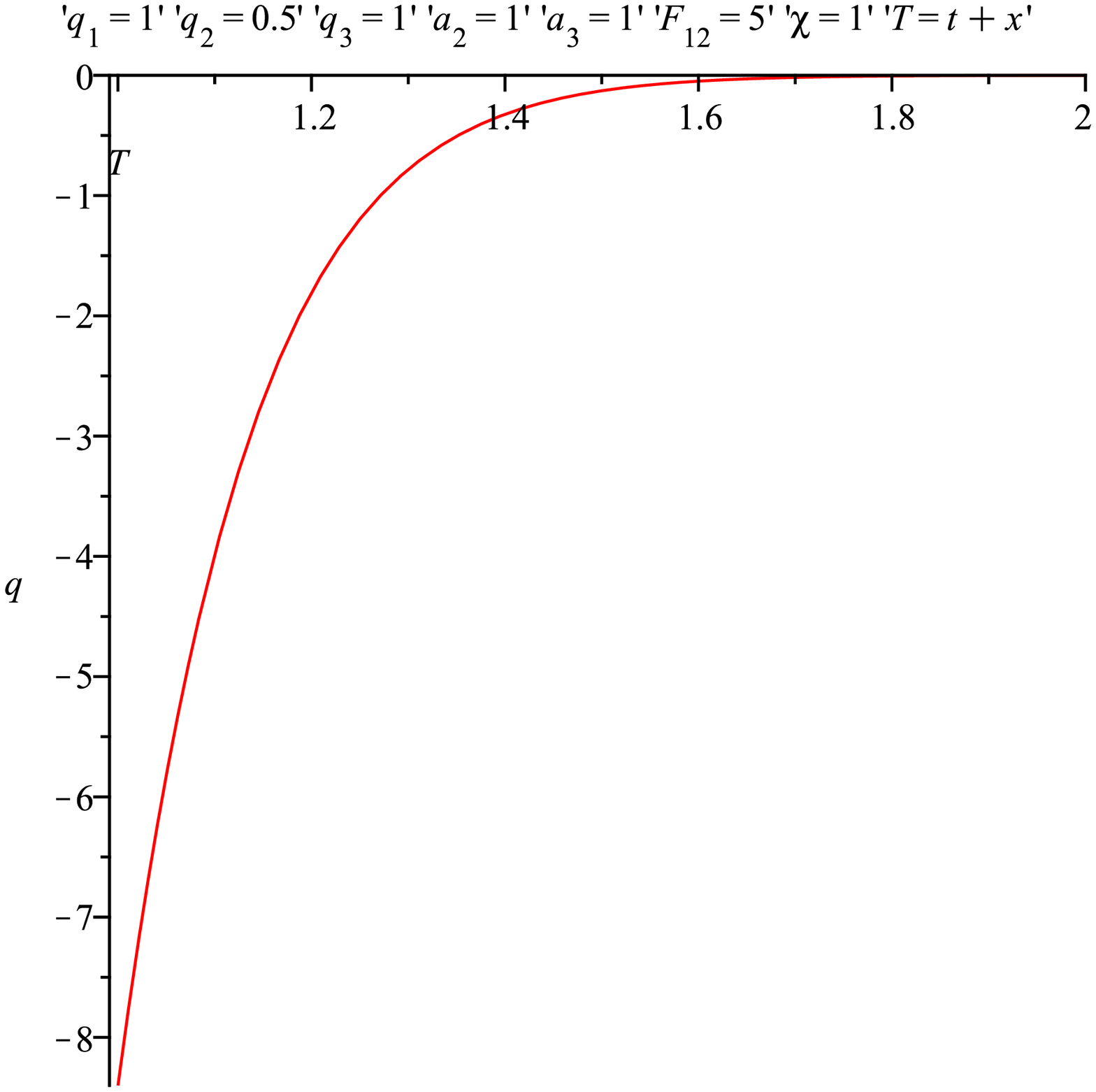} \\
\end{tabular}
\caption{ (\textit{Left})  The variation of the volume of the
Universe with respect to time.. (\textit{Middle}) The variation of
$\frac{\sigma}{\theta}$ with respect to time. (\textit{Right}) The
deceleration parameter is shown with respect to time.}
\end{figure*}
\begin{figure*}[thbp]
\begin{tabular}{rl}
\includegraphics[width=4.5cm]{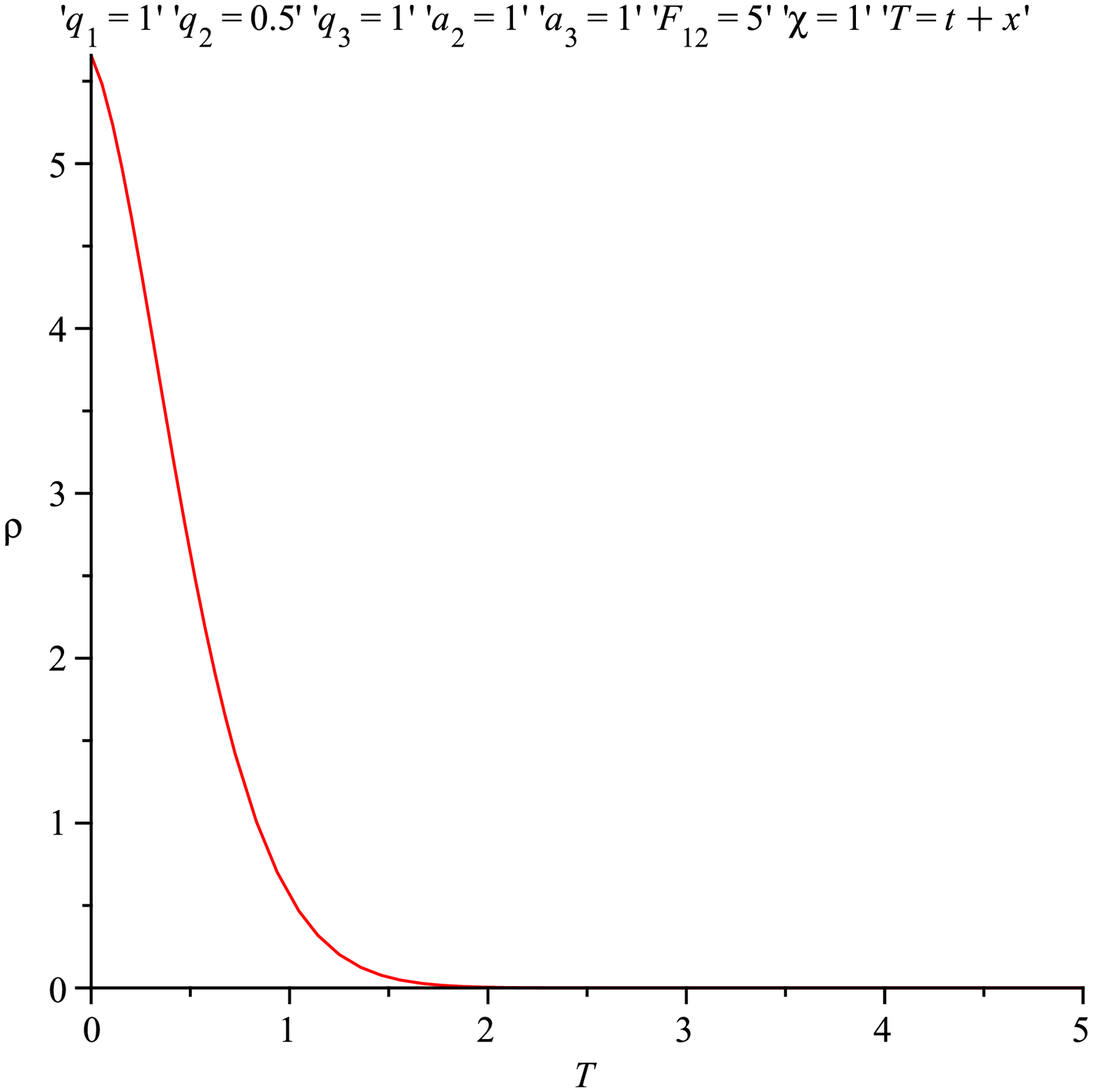}&
\includegraphics[width=4.5cm]{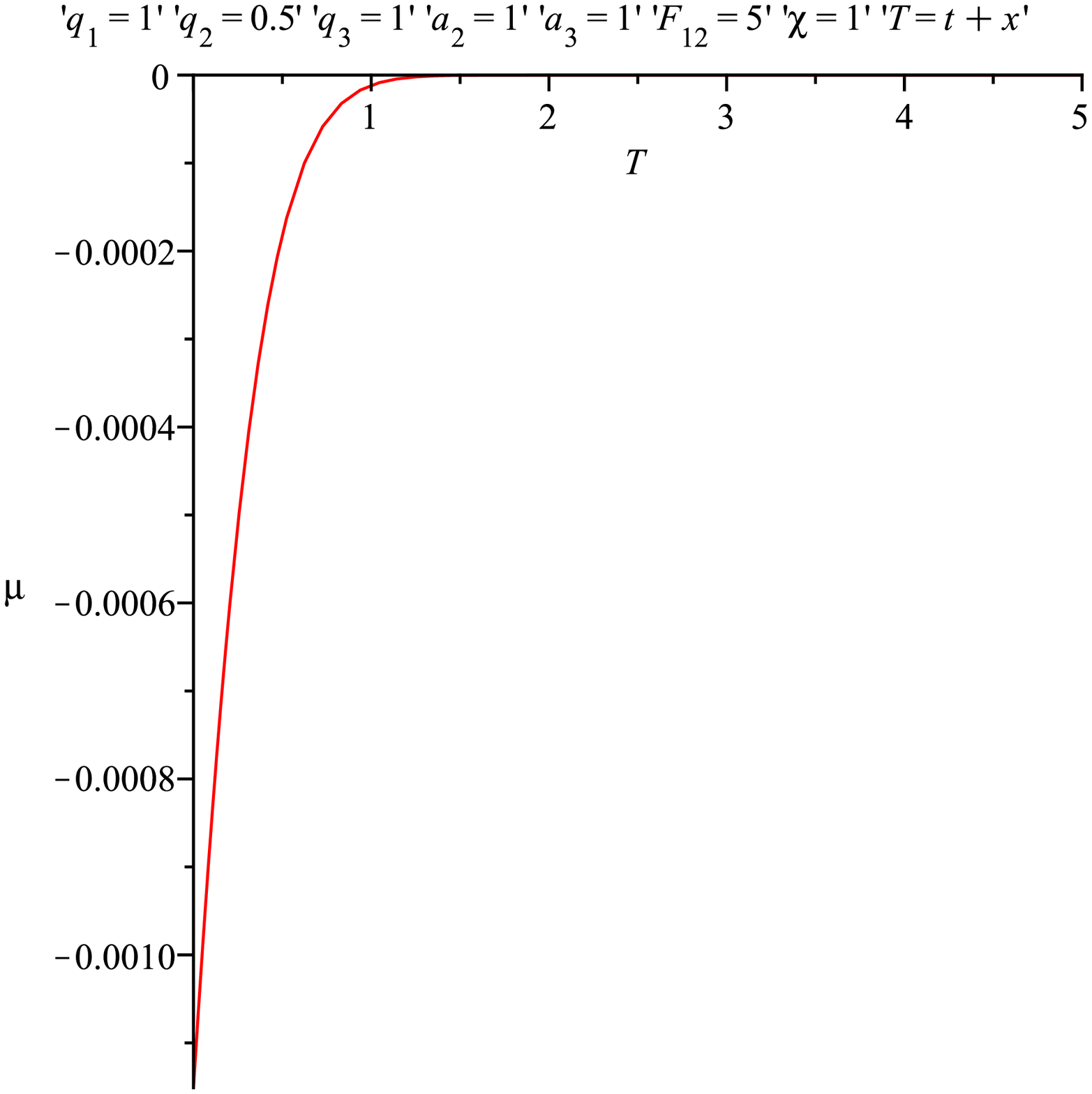}
\includegraphics[width=4.5cm]{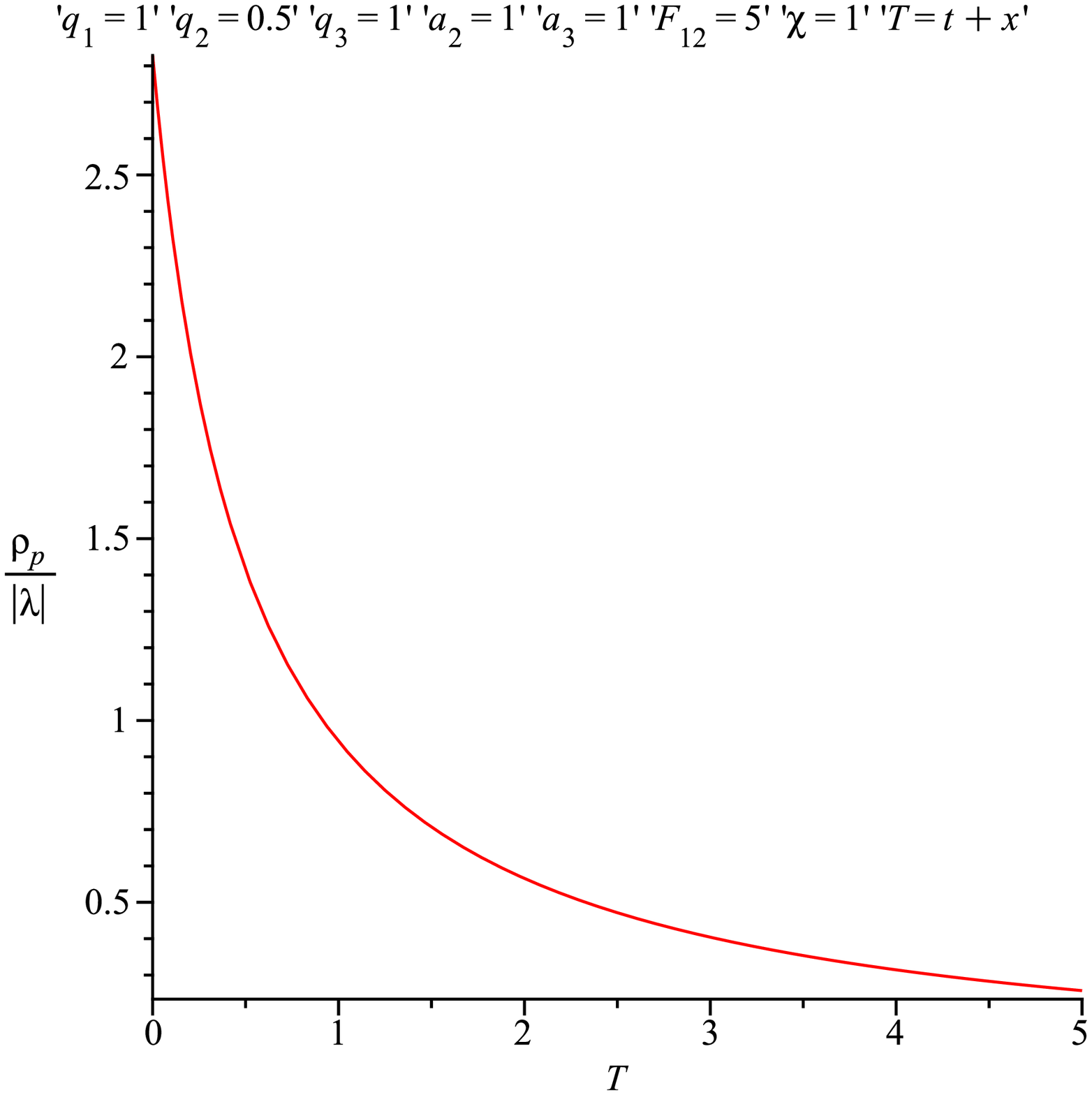} \\
\end{tabular}
\caption{ (\textit{Left})  The variation of energy density with
respect to time. (\textit{Middle}) The variation of magnetic
Permeability with respect to time. (\textit{Right}) The variation of
particle density with respect to time.}
\end{figure*}


\end{document}